\documentclass{article}

\usepackage{amsmath}
\usepackage{amssymb}
\usepackage{amstext}
\usepackage[all]{xy}
\usepackage{mathrsfs}
\usepackage{pstricks}
\usepackage{pst-3d}
\usepackage{pst-node}
\usepackage{pst-fill}
\usepackage{pst-coil}
\usepackage{wrapfig}
\usepackage{graphicx}

\renewcommand{\arraystretch}{1.7}

\def\={=}                               
\def\df{\frac{}{}}                              

\renewcommand{\mathbf}[1]{\boldsymbol{#1}}

\def\w{\wedge}                                  
\def\wt#1{\widetilde{#1}}                       
\def\pdiff#1#2{\frac{\partial#1}{\partial#2}}   
\def\SOdiff#1#2{\frac{d^{2}#1}{d#2^{2}}}        
\def\({\left(}                                  
\def\){\right)}                                 
\def\[{\left[}                                  
\def\]{\right]}                                 
\def\conj#1{\overline{#1}}                      
\def\tensor{\otimes}
\def\Re{\text{Re}}                              

\def\qquadand{\qquad\text{and}\qquad}

\def\tm#1{{\mbox{\tiny $#1$}}}
\def\tt#1{{\mbox{\tiny #1}}}

\def\man#1{{\cal #1}}
\def\real{\mathbb{R}}
\def\Lf{\mathscr{L}}

\def\ep#1{\epsilon_{#1}}
\def\g{\underline{g}}
\def\ginv{g^{\mbox{\tiny $-1$}}}
\def\cc{c}

\def\wh#1{\widehat{#1}}
\def\omegaa{\omega,\man{L}_{\nu}}

\def\ee{{\mathbf e}}
\def\bb{{\mathbf b}}
\def\dd{{\mathbf d}}
\def\hh{{\mathbf h}}

\def\BB{{\mathbf B}}
\def\DD{{\mathbf D}}

\def\A{\mathbb{A}}

\def\pphi{{\Phi}}
\def\ppsi{{\Psi}}
\def\hpsi{\wh{\psi  }}

\def\k{\kappa}
\def\kr{\man{K}}

\def\TE{\mbox{\tiny TE},\nu}
\def\TM{\mbox{\tiny TM},\nu}
\def\TTE{\mbox{\tiny TE}}
\def\TTM{\mbox{\tiny TM}}
\def\SSigma{\mbox{\tiny $\Sigma$}}

\def\L{\mbox{\tiny L}}
\def\E{\mbox{\tiny E}}
\def\O{\mbox{\tiny O}}
\def\kx{k_{x}}
\def\ky{k_{y}}
\def\kz{k_{z}}
\def\kk{\mathbf{k}}
\def\nx{n_{x}}
\def\ny{n_{y}}

\def\Lx{L_{x}}
\def\Ly{L_{y}}
\def\Lz{L_{z}}
\def\Lf{L_{f}}

\def\rrr{ x,y,z }
\def\chiI{\chi_{\I}}
\def\chiII{\chi_{\II}}
\def\chiIII{\chi_{\III}}
\def\chinu{\chi_{\nu}}
\def\zetanu{\zeta_{\nu}}
\def\zetaI{\zeta_{\I}}
\def\zetaII{\zeta_{\II}}
\def\zetaIII{\zeta_{\III}}

\def\I{\mbox{\tiny I}}
\def\II{\mbox{\tiny II}}
\def\III{\mbox{\tiny III}}
\def\reg{\mbox{\tiny I,\,II,\,III}}
\newcommand{\MD}[6]{  \ensuremath{   {#1}^{{#2},{#3}}_{{#4}, {\cal {#5}}} #6   }     }
\newcommand{\MMD}[6]{  \ensuremath{   {#1}^{{#2},{#3}}_{{#4}, { {#5}}} #6   }     }

\def\inv{\mbox{\tiny $-1$}}

\def\f{ \mathcal{ G }  }

\begin{document}
\title{On the computation of Casimir stresses in open media and Lifshitz theory}
\author{\vspace{-0.2cm} Shin-itiro Goto$^{1,2}$, Robin W Tucker$^{1,2}$ and Timothy J Walton$^{1,2}$ \\
\begin{tabular}{rl}
    \vspace{-0.3cm} $^{1}$ & \hspace{-0.5cm} \mbox{\footnotesize  Department of Physics, Lancaster University, Lancaster, LA1 4YB} \\
    $^{2}$ & \hspace{-0.5cm} \mbox{\footnotesize  The Cockcroft Institute, Daresbury Laboratory, Warrington, WA4 4AD}
\end{tabular} \\
}
\maketitle

\begin{abstract}
    A classification of the electromagnetic modes on open and closed spatial domains containing media with piecewise homogeneous permittivities is used to facilitate the derivation of quantum induced Casimir stresses in dielectrics. By directly exploiting the complex analytic properties of solutions of the macroscopic Maxwell equations for open systems it is shown how regular expressions for such stresses can be expressed in terms of double integrals involving either real or pure imaginary frequencies associated with harmonic modes in conformity with the Lifshitz theory for separated planar dielectric half-spaces. The derivation is self-contained without recourse to the Krein formula for a density of states or mode regularization and offers a more direct approach to other open systems. \\
    \quad \\
    PACS numbers: 03.70.+k, 12.20.Ds, 31.30.J-, 42.50.Ct, 42.50.Pq
\end{abstract}

\section{Introduction}

The problem of calculating the mechanical stresses induced by quantum electromagnetic fluctuations in a material medium continues to attract
attention in the scientific literature. Aside from its purely academic interest, its resolution has wide implications in many areas of science and technology on many disparate scales \cite{Bordag,Boyer,Wolfram,Milton2,Milton3,Pendry,Lifshitz1,Lifshitz2,Lifshitz3}. These vary from quantum optics, meta-materials, nanofluidics, nanomachinery, cold atom technology and early Universe cosmology. Lifshitz \cite{Lifshitz2} was one of the first to tackle the problem in the context of an open dielectric medium containing a planar layer of free space. Based on the fluctuation-dissipation theorem, the intricate methodology employed by this author has never been fully appreciated and alternative, less opaque methods have been explored in this context. Suffice to say that there remains no general consensus on how best to extend these methods to less idealized situations of more direct relevance to applied science.

The general problem involves a number of subtleties of varying degrees of complexity. While most of these are fully appreciated by experts in the field, to our knowledge, a general treatment outlining theses subtleties is not in evidence. Indeed, without establishing a well  posed problem in a mathematical sense, the viability of any  testable prediction becomes difficult. In general, to predict the significance of Casimir type stresses one is confronted with establishing a quantization programme of the Maxwell field in a medium at rest with possible anisotropic and inhomogeneous specified constitutive properties \cite{Leonhardt1,Leonhardt2,goto2012numerical,Philbin,Philbin1}. The medium response to the macroscopic electromagnetic fields may depend non-locally on time and space (temporal and spatial dispersion) in a linear or non-linear manner and induce thermal effects that also require constitutive properties for their inclusion into the programme. Thus, even at a local level, one is confronted with the formulation of a potentially difficult thermo-electrodynamic problem. For a stationary medium (in time and space) such a formulation also requires a specification of the support of the medium in space and, in particular, whether it is open or closed and the behaviour of electromagnetic fields across any interfaces between media with different constitutive properties. For example, at dielectric interfaces one expects certain components of the electromagnetic fields to induce surface charge densities and at metallic boundaries one expects both induced surface charge densities and surface current densities.

Casimir stresses for media at rest are {\it stationary} phenomena arising from the quantum expectation value of components of a (Cauchy) stress operator on a (possibly rigged) Hilbert sub-space of states representing the quantum electromagnetic fields in the medium. Such states must therefore arise as a complete eigen-set of some symmetric operator with \textit{real} eigen-values. For certain problems with piecewise homogeneous media and some degree of spatial symmetry certain components of the stress may be obtained by appeal to conservation laws by considering the work done in dilating the material. This inevitably
requires the computation of a pressure density at an interface from an energy density. The computation of a regularized energy density quantum
expectation value has proved an effective route for estimating Casimir stresses \cite{Reuter,GTW,Visser}. A common approach leading to such a regularized energy density exploits the analytic properties of certain spectrum generating functions in  the complex frequency plane. Such functions are chosen to reproduce the real spectra associated with all the electromagnetic eigen-modes allowed in the medium. For media confined in finite volumes of space the allowed modes are those that satisfy the assumed boundary conditions on the macroscopic electromagnetic fields on any bounding surfaces as well as at possible interfaces between media with piecewise disjoint constitutive properties. For  media in closed domains, the spectrum consists of infinitely many discrete spectral values of some operator. If the media are wholly or partially unconfined and the spectrum is continuous then determining the density of allowed states is, in general, a difficult problem.

It should be stressed that any theoretical prediction based on constitutive models that neglect dissipation are unlikely to yield accurate experimental predictions since dissipation accompanies most physical processes. It remains unclear whether the ``fluctuation-dissipation theorem'', on which Lifshitz based his theory, is adequate to accommodate all physical processes. However, the focus of our article offers an alternative description that promotes complex analyticity at its heart in place of this ``theorem''. Given that dispersion and dissipation are intimately related for causal processes (via the Kramers-Kronig relations), we feel that a reformulation that stresses complex analyticity, even in the absence of dissipation, has some merit for future developments, for example, along the lines promoted in \cite{intravaia2008casimir, intravaia2009casimir}.

Since the early work by Lifshitz and others on the simple open planar layered dielectric there has been much debate on whether proposed formulae for Casimir stresses in media properly account for the full spectrum of allowed states. While this work was in progress two papers appeared dealing with similar aspects of Lifshitz theory. Each approached the problem of making explicit the derivation of the Lifshitz formula for the interaction between planar dielectric half-spaces separated by a vacuum slab of finite thickness from somewhat different perspectives.  In \cite{nesterenko2012lifshitz}, emphasis was placed on evaluating a mode summation for such a configuration by appealing to a formula of Krein \cite{Krein} derived in the context of scattering theory. Its derivation drew heavily on the language of this theory in 1-space dimension with (quantized) discrete bound states and continuous scattering states despite the fact that in 3-dimensional open systems all allowed states that contribute to a Casimir stress in such a piecewise homogeneous configuration have continuous spectra.  In \cite{bordag2012electromagnetic}, emphasis was placed on a regularized spectral spectral summation based on a density of states derived in the context of a confined configuration whose volume tended to infinity.  The subsequent summation for an open system used integration variables that necessitated a careful regularization and renormalization process in order to relate the Lifshitz integral to a finite mode summation.

We believe that the approach adopted in this article offers a more succinct and self-contained method to unify a number of aspects in these and other papers that have attempted to penetrate the Lifshitz theory. Its point of departure is a classification of solutions to the macroscopic Maxwell equations for classical fields in a piecewise continuous medium subject to both boundary conditions for confined media and bounded-ness conditions at all points in space for  confined and unconfined media. These conditions are distinguished from any discontinuity conditions that have to be satisfied on all interfaces where constitutive properties of the media may change abruptly. The general approach is established in section~\ref{Sect:Maxwell} and made explicit for fields confined to a perfectly conducting cuboid in section~\ref{CavitySect}. A state labelling scheme is defined in section~\ref{Sect:SpectraCavity} and a complete set of modes in the cuboid described in terms of a real frequency spectrum is summarized in \ref{BoundAppendix}. In section~\ref{Sect:DispOpen} and \ref{OpenAppendix} this is contrasted with a description of a complete set of states that form a basis for describing electromagnetic modes in open media in planar half-spaces separated by a finite width planar vacuum slab. It is shown that these modes have real angular frequencies arising as roots of expressions that can all be related to complex amplitudes by a simple analytic continuation in the complex angular frequency plane.  By exploiting this complex analytic structure the Lifshitz formula in terms of pure imaginary frequencies becomes related to a real double integral over functions involving real frequencies and arises as an identity from the Cauchy integral formula. This double integral has an immediate interpretation in terms of a continuous mode summation over allowed real frequencies associated with allowed field configurations in the open system without recourse to input from the Krein formula and for media with physically relevant constitutive properties is finite without recourse to regularization or renormalization.

The identity is imbued with physical content by exploiting a freedom to normalize the continuum states in a distributional context. If one demands that the Lifshitz expression should describe the attraction between perfectly conducting plates when the two half spaces emulate perfect conductors one can then derive physical quantum induced stresses in the planar half-spaces for arbitrary (physical) dielectric media. However, throughout this paper, attention is restricted to dissipation-free, thermally inactive models for dielectric media. The advantages of this approach in contemplating the computation of such stresses in  more general geometrical configurations and media with more complex constitutive properties are discussed in the concluding section.

\section{The Macroscopic Maxwell Equations in Media}\label{Sect:Maxwell}
An essential precursor in this programme is the enumeration of a set of classical stationary modes defined as solutions to the Maxwell system on a prescribed domain of $\real^{3}$ and subject to specified conditions on any boundaries and at possible regions in the domain where the constitutive properties change discontinuously. Before prescribing particular constitutive models and boundary conditions, consider the {\it local} problem of solving the macroscopic Maxwell equations in
the absence of free charge and current:
\begin{eqnarray}\label{CovMaxEq}
    dF &=& 0 \qquadand d\star G \= 0
\end{eqnarray}
where $F,G$ are 2-forms on space-time and $\star$ denotes the Hodge map associated with the Minkowski metric tensor
\begin{eqnarray*}
    g &=& -\cc^{2} dt \tensor dt + \g.
\end{eqnarray*}
Here $\g$ is the Euclidean metric tensor on $\real^{3}$ and $t$ denotes local laboratory time. The decomposition of the 2-forms $F$ and $G$ into
time-dependent 1-form fields $\{\ee,\bb,\dd,\hh\}$ in an inertial $g$-orthonormal co-frame $\{\cc\,dt, \, e^{1},\, e^{2}, \, e^{3}\}$ on some domain $U\subset\real^{3}$ is effected by writing
\begin{equation}\label{FGdef}
\begin{split}
     F &\= \ee \w \wt{V} - \star\(\cc\bb \w \wt{V}\) \\
     G &\= \dd \w \wt{V} - \star\(\frac{\hh}{\cc} \w \wt{V}\)
\end{split}
\end{equation}
where $\wt{V}=-\cc\,dt$ with $\cc$ the vacuum speed of light. Then (\ref{CovMaxEq}) yields the exterior system:
\begin{eqnarray}\label{MaxEq}
\begin{tabular}{lllll}
    $d\ee $ & $\hspace{-0.35cm} \= -\partial_{t}\BB$, & \quad & $d\hh$ & $\hspace{-0.35cm} \= \partial_{t}\DD$ \\
    $d\BB $ & $\hspace{-0.35cm} \= 0$, & \quad & $d\DD$ & $\hspace{-0.35cm} \=0$.
\end{tabular}
\end{eqnarray}
where $\BB=\#\bb$ and $\DD=\#\dd$ in terms of the Hodge map $\#$ on $\real^{3}$ defined by the relation:
\begin{eqnarray*}
    \star 1 &=& \wt{V} \w \# 1
\end{eqnarray*}
with $\#1 = e^{1} \w e^{2} \w e^{3}$.  On any part of a common interface $\Sigma$ between regions $U_{\I}$ and $U_{\II}$ with different constitutive properties, the fields $\ee$, $\bb$, $\dd$ and $\hh$ have some components that may be discontinuous. Their discontinuities follow from extending the local system (\ref{MaxEq}) to a domain where it can accommodate  discontinuous solutions. If the interface $\Sigma$ is part of some stationary hypersurface in $\real^{3}$ given by the equation $f=0$ where $df\neq 0$, the field discontinuities, denoted with $\[\phantom{m}\]$, satisfy:
\begin{eqnarray*}
\begin{tabular}{lllll}
    $\displaystyle \left. \df \[\ee\] \w df \right|_{\Sigma}$ & $\hspace{-0.35cm} \= 0$, & \quad & $\displaystyle \left. \df \[\BB\] \w df \right|_{\Sigma}$ & $\hspace{-0.35cm} \= 0$, \\
    $\displaystyle\left. \df \[\DD\] \w df \right|_{\Sigma}$ & $\displaystyle \hspace{-0.35cm} \= \left. \df \sigma \w df\right|_{\Sigma}$,& \quad & $\displaystyle \left. \df \[\hh\] \w df \right|_{\Sigma}$ & $\displaystyle \hspace{-0.35cm} \= \left. \df \man{J} \w df \right|_{\Sigma}$
\end{tabular}
\end{eqnarray*}
for some induced surface charge 2-form $\sigma$ and induced surface current 1-form $\man{J}$. If $W$ is any time-dependent 1-form on $\real^{3}$, call $t_{\SSigma}W = W\w df$ its associated tangential 2-form with respect to $\Sigma$ and $n_{\SSigma}W=(\#W )\w df$ the associated normal 3-form
with respect to $\Sigma$. Hence the above interface conditions may be written:
\begin{eqnarray*}
\begin{tabular}{lllll}
    $\displaystyle \left. \df \[t_{\SSigma}\ee\] \right|_{\Sigma}$ & $\hspace{-0.35cm}\= 0$, & \quad & $\displaystyle \left. \df \[n_{\SSigma}\bb\] \right|_{\Sigma}$ & $\hspace{-0.35cm}\= 0$, \\
    $\displaystyle\left. \df \[n_{\SSigma}\dd\] \right|_{\Sigma}$ & $\displaystyle \hspace{-0.35cm}\= \left. \df \sigma \w df\right|_{\Sigma}$,& \quad & $\displaystyle \left. \df \[t_{\SSigma}\hh\] \right|_{\Sigma}$ & $\displaystyle \hspace{-0.35cm} \= \left. \df \man{J} \w df \right|_{\Sigma}$.
\end{tabular}
\end{eqnarray*}
If $U_{\I}$ is a perfect conductor, all fields in $U_{\I}$ are zero and so on $\Sigma$:
\begin{eqnarray*}
\begin{tabular}{lllll}
    $\displaystyle \left. \df t_{\SSigma}\ee \right|_{\Sigma}$ & $\hspace{-0.35cm} \= 0$, & \quad & $\displaystyle \left. \df n_{\SSigma}\bb \right|_{\Sigma}$ & $\hspace{-0.35cm} \= 0$, \\
    $\displaystyle\left. \df n_{\SSigma}\dd \right|_{\Sigma}$ & $\displaystyle \hspace{-0.35cm} \= \left. \df \sigma \w df\right|_{\Sigma}$,& \quad & $\displaystyle \left. \df t_{\SSigma}\hh \right|_{\Sigma}$ & $\displaystyle \hspace{-0.35cm} \= \left. \df \man{J} \w df \right|_{\Sigma}$.
\end{tabular}
\end{eqnarray*}
If $U_{\I}$ and $U_{\II}$ are perfect insulators, one cannot sustain surface currents on $\Sigma$ (although induced surface charges may arise) and so on $\Sigma$
\begin{eqnarray*}
\begin{tabular}{lllll}
    $\displaystyle \left. \df \[t_{\SSigma}\ee\] \right|_{\Sigma}$ & $\hspace{-0.35cm} \= 0$, & \quad & $\displaystyle \left. \df \[n_{\SSigma}\bb \] \right|_{\Sigma}$ & $\hspace{-0.35cm} \= 0$, \\
    $\displaystyle\left. \df \[ n_{\SSigma}\dd \] \right|_{\Sigma}$ & $\displaystyle \hspace{-0.35cm} \= \left. \df \sigma \w df\right|_{\Sigma}$,& \quad & $\displaystyle \left. \df \[ t_{\SSigma}\hh \] \right|_{\Sigma}$ & $\displaystyle \hspace{-0.35cm} \= 0$.
\end{tabular}
\end{eqnarray*}
In the former case, $\sigma$ and $\man{J}$ are not in general specified a-priori. In the latter case, $\sigma$ is not in general specified. Note that in terms of the pull-back $\Sigma^{*}W$ of $W$ to $\Sigma$ one has
\begin{eqnarray*}
    \left. \df i_{\partial_{f}}t_{\SSigma}W\right|_{\Sigma} &=& -\Sigma^{*}W
\end{eqnarray*}
where $i_{\partial_{f}}$ denotes the interior operator. This has implications for the time-harmonic Maxwell system. Then for any real, time-harmonic 1-form $W$ on $\real^{3}$, define the complex 1-form  $W_{\omega}$ on $\real^{3}$ by
\begin{eqnarray*}
    W &=& \Re\( W_{\omega}e^{-i\omega t} \)
\end{eqnarray*}
with $\omega\in\real^{+}$ and the complex conjugate $\conj{W}_{\omega}=W_{-\omega}$. The time harmonic Maxwell system for
$\{\ee_{\omega},\bb_{\omega},\dd_{\omega},\hh_{\omega}\}$ is
\begin{eqnarray*}
\begin{tabular}{lllll}
    $d\ee_{\omega}$ & $\hspace{-0.35cm} \= i\omega\BB_{\omega}$, & \qquad & $d\hh_{\omega}$ & $\hspace{-0.35cm} \= -i\omega\DD_{\omega}$ \\
    $d\BB_{\omega}$ & $\hspace{-0.35cm} \= 0$, & \qquad & $d\DD_{\omega}$ & $\hspace{-0.35cm} \=0$.
\end{tabular}
\end{eqnarray*}
and the above interface conditions are applied to each harmonic component. The condition $\left.\df\[t_{\SSigma}\ee_{\omega}\]\right|_{\Sigma}=0$ is
equivalent to $\[\Sigma^{*}\ee_{\omega}\]=0$. Furthermore, since $d\ee_{\omega}=i\omega\BB_{\omega}$ and $d\Sigma^{*}=\Sigma^{*}d$, one has, with $\omega \neq 0$,  $\[\Sigma^{*}\BB_{\omega}\]=0$ or $\left.\df\[n_{\SSigma}\bb_{\omega}\]\right|_{\Sigma}=0$. Similarly, if
$\left.\df\[n_{\SSigma}\bb_{\omega}\]\right|_{\Sigma}=0$, one has $\left.\df\[t_{\SSigma}\ee_{\omega}\]\right|_{\Sigma}=0$. Then for an interface with a perfect conductor, one can impose $\left.\df t_{\SSigma}\ee_{\omega}\right|_{\Sigma}=0$ (or $\left.\df
n_{\SSigma}\bb_{\omega}\right|_{\Sigma}=0$) and calculate $\sigma_{\omega}$ and $\man{J}_{\omega}$ from
$\left.\df\[n_{\SSigma}\dd_{\omega}\]\right|_{\Sigma}$ and $\left.\df\[t_{\SSigma}\hh_{\omega}\]\right|_{\Sigma}$ respectively. Similarly, for an interface between two perfect insulators, one can impose $\left.\df\[t_{\SSigma}\hh_{\omega}\]\right|_{\Sigma}=0$ and
$\left.\df\[t_{\SSigma}\ee_{\omega}\]\right|_{\Sigma}=0$ (or $\left.\df\[t_{\SSigma}\hh_{\omega}\]\right|_{\Sigma}=0$ and
$\left.\df\[n_{\SSigma}\bb_{\omega}\]\right|_{\Sigma}=0$) and then calculate $\sigma_{\omega}$ from
$\left.\df\[n_{\SSigma}\dd_{\omega}\]\right|_{\Sigma}$. Note that these  conditions have general applicability to any stationary curved or planar interface that coincides with a segment of the locus $f=0$ in $\real^{3}$ where $df \neq 0$. A unit normal field to the surface is the vector field $N_{f}$ on $\real^{3}$ in the vicinity of $\Sigma$  defined by
\begin{eqnarray*}
    N_{f} &\equiv& \frac{\ginv(df,-)}{\sqrt{\ginv(df,df)}}.
\end{eqnarray*}

\section{The Cavity Mode Problem}\label{CavitySect}
It will prove valuable to contrast the spectral content of solutions to the above Maxwell system describing electromagnetic modes in media in both closed and open domains of space. In this section, attention is restricted to fields confined to a {\it closed} region of $\real^{3}$ that does not change with time. Typically, this is achieved by confining the fields in a closed cavity. Such a cavity may be considered as a void in a larger domain composed of a medium that cannot support non-zero fields or a region surrounded by a surface composed of a perfect conductor. The cavity will be supposed filled with a piecewise continuous, piecewise homogeneous medium with constant permeability and real permittivity (that may depend on the frequency of harmonic fields). Since the permittivity is assumed real, all losses due to dispersion are assumed negligible and thermal effects ignored.

The equation $dF=0$ is locally satisfied by $F=dA$ for some real 1-form $A$ on any open space-time domain (with compact space-like sections) in the history of the cavity. Writing $A=\cc\,\phi \, dt - \cc\A$ in terms of a time-dependent 0-form $\phi$ and time-dependent 1-form $\A$ on  any such section $U\subset\real^{3}$, one has from
(\ref{FGdef})
\begin{eqnarray*}
    \ee &=& -\partial_{t}\A - d\phi \qquadand \bb \= \#d\A.
\end{eqnarray*}
In a simple dielectric medium at rest:
\begin{eqnarray}\label{CREL}
    \dd_{\omega} &=& \ep{0}\k\ee_{\omega} \qquadand \hh_{\omega} \= \mu_{0}^{\inv}\bb_{\omega}
\end{eqnarray}
for a real permittivity $\ep{0}\k$ that may depend on $\omega$.
In a gauge with $d\#\(\ep{0}\k\A_{\omega}\)=0$ and $d\phi_{\omega}=0$, the system (\ref{CovMaxEq}) reduces to
\begin{eqnarray*}
     \#d\#d\A_{\omega} - \frac{\k\omega^{2}}{\cc^{2}}\A_{\omega} &=& 0
\end{eqnarray*}
subject to
\begin{eqnarray}\label{GaugeCond}
    d\#\(\k\A_{\omega}\) &=& 0
\end{eqnarray}
on $U\subset\real^{3}$. Henceforth assume that $U=\bigcup_{\nu}U_{\nu}$ is a {\it simply connected} sub-domain of $\real^{3}$ and exploit the fact that any 1-form can then be decomposed in terms of forms in the kernel of $d$ and $\#d\#$ (Hodge-Weyl-Friedrichs decomposition). In the Maxwell context, such forms are often designated by the labels TE and TM. Thus one may write
\begin{eqnarray*}
    \A_{\omega}^{\nu} &=& \A_{\omega}^{\TE} + \A_{\omega}^{\TM}
\end{eqnarray*}
where
\begin{eqnarray}\label{ATETMsplit}
    \A_{\omega}^{\TE} &\equiv& \#d\phi^{\TE}_{\omega} \qquadand \A_{\omega}^{\TM} \equiv \#d\#d\phi^{\TM}_{\omega}
\end{eqnarray}
in terms of the complex {\it pre-potential 1-forms} $\phi^{\TE}_{\omega},\phi^{\TM}_{\omega}$.  General solutions for all the Maxwell fields can then be determined in terms of the $\phi^{s,\nu}_{\omega}$ for $s\in\{\text{TE,TM}\}$, obtained by substituting (\ref{ATETMsplit}) into
\begin{eqnarray}\label{HelmMaxDom}
    \#d\#d\A^{s,\nu}_{\omega} - \frac{\k^{\nu}\omega^{2}}{\cc^{2}}\A^{s,\nu}_{\omega} &=& 0 \qquad \qquad \nu \= \text{I, II,} \ldots
\end{eqnarray}
Since $\#\#=1$ and $d^{2}=0$, the 1-forms $\A^{\TE}_{\omega},\A^{\TM}_{\omega}$ automatically satisfy the gauge condition
$d\#\(\k^{\nu}\A^{s,\nu}_{\omega}\) = 0$. For spatially homogeneous media the gauge conditions are independent of sub-domain. From the above
definitions, it is apparent that the components of the $\phi^{\TE}_{\omega}$ should be at least twice differentiable on $U_{\nu}$ and the components of $\phi^{\TM}_{\omega}$ three times differentiable on $U_{\nu}$.

The general solutions in each sub-domain can be used to define gauge-invariant electric and magnetic fields that satisfy the above interface
conditions across the interfaces between distinct sub-domains as well as  physically motivated boundary conditions. With such (homogeneous)
conditions imposed such solutions should determine a complete set of cavity modes with spectral parameter $\omega^s$. For perfectly conducting
boundary conditions (and non-absorbing media at zero temperature) the angular frequencies $\omega^s$  of these modes will be real and since $U$ is assumed to have finite volume with boundary each mode will be associated with  a possible denumerably infinite set of discrete values. If
$\phi^{s,\nu}_{\omega}$  defined by (\ref{ATETMsplit})  yields $\A_{\omega}^{s,\nu}$ satisfying (\ref{HelmMaxDom}) on $U_{\nu}$, a solution
$\Phi^{s}_{\omega}$ on $U$ is then written $\sum_{\nu}\phi^{s,\nu}_{\omega}$ where the summation is over all sub-domains where the permittivity is smooth. The corresponding gauge-invariant electric and magnetic  fields in the cavity will be written  $\ee^{s}_{\omega}
=\sum_{\nu}\ee^{s,\nu}_{\omega}$  and $\bb^{s}_{\omega}= \sum_{\nu}\bb^{s,\nu}_{\omega}$ respectively.
The determination of a complete set of such modes yields a basis for solving the initial value (Cauchy) problem in the cavity and is in general a non-trivial problem. In the following attention will be restricted to stationary modes described by real spectra  and cavities where the TE and TM modes decouple. The completeness relations can  be expressed in terms of a choice of an orthogonal basis for a suitable Hilbert space of 1-forms over $U$. An appropriate inner product, with  $(\pphi,\pphi)_{\tm{U}} >0$, is defined by\footnote{The 1-form $\pphi$ (with complex conjugate $\conj{\pphi}$) should have co-ordinate components that are  continuous  on each sub-domain $U_{\nu}$ with $(\pphi,\pphi)_{\mbox{\tiny U}}$ finite.}
\begin{eqnarray}\label{IPROD}
    (\pphi,\ppsi)_{\tm{U}} &\equiv& \int_{U} \conj{\pphi} \w \# \ppsi .
\end{eqnarray}
It follows that
\def\BBOX{ \#d\#d }
\begin{eqnarray*}
    (\BBOX\pphi,\ppsi)_{\mbox{\tiny $U$}} &=& (\pphi,\BBOX\ppsi)_{\mbox{\tiny $U$}} + \int_{\partial U}\( \conj{\pphi} \w \# d\ppsi - \ppsi \w
    \#d\conj{\pphi} \).
\end{eqnarray*}
Writing $\pphi=\A^{s}_{\omega},\,\,\ppsi=\A^{s}_{\omega'}$  so that $\ee^{s}_{\omega}=i\omega\A^{s}_{\omega}, \bb^{s}_{\omega}=\#d\A^{s}_{\omega}$, the operator $ \#d\#d  $ is symmetric with respect to $({\,\,},{\,\,})_{\tm{U}}$ provided
\begin{eqnarray}\label{HilSpCond}
    (\partial U)^{*}\( \ee^{s}_{\omega} \w \bb^{s}_{\omega'} - \ee^{s}_{\omega'} \w \bb^{s}_{\omega} \) &=& 0  \qquad \forall s, \omega,\omega' .
\end{eqnarray}
This condition is satisfied for fields satisfying the perfectly conducting boundary conditions $(\partial U)^{*}\ee^{s}_{\omega}=0$ or $(\partial U)^{*}\hh^{s}_{\omega}=0$ discussed above.

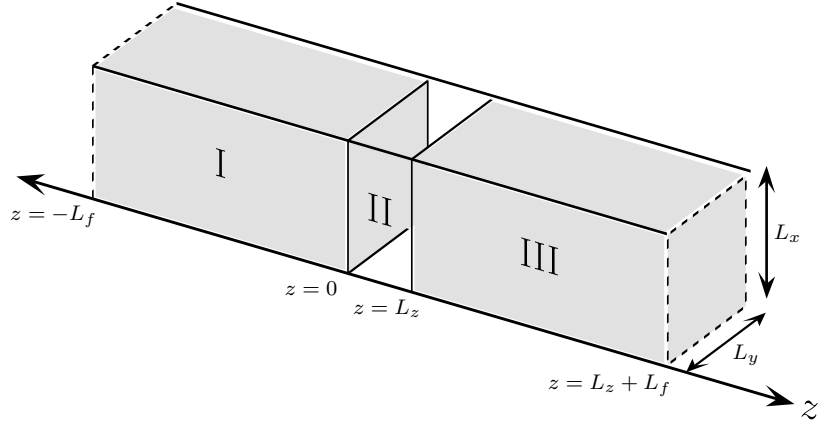
\begin{figure}[h!]
\begin{center}
\setlength{\unitlength}{1cm}
\begin{picture}(10,7)
\psset{viewpoint=1 -1.6 1}
    \definecolor{LightGrey}{rgb}{0.88,0.88,0.88}
    \ThreeDput[normal=0 1 0](11,0,4){\psframe[linewidth=1.2pt, fillstyle=solid, fillcolor=LightGrey, linecolor=white](0,0)(4,2)}
    \ThreeDput[normal=0 0 1](7,0,6){\psframe[linewidth=1.2pt, fillstyle=solid, fillcolor=LightGrey, linecolor=white](0,0)(4,2)}
    \ThreeDput[normal=0 1 0](6,0,4){\psframe[linewidth=1.2pt, fillstyle=solid, fillcolor=LightGrey, linecolor=white](0,0)(4,2)}
    \ThreeDput[normal=0 0 1](2,0,6){\psframe[linewidth=1.2pt, fillstyle=solid, fillcolor=LightGrey, linecolor=white](0,0)(4,2)}
    \ThreeDput[normal=1 0 0](6,0,4){\pspolygon[linewidth=1.2pt, fillstyle=solid, fillcolor=LightGrey, linecolor=LightGrey](0,0)(2,0)(2,2)(0,2)}
    \ThreeDput[normal=1 0 0](11,0,4){\psframe[linewidth=1.2pt, fillstyle=solid, fillcolor=LightGrey, linestyle=dashed, dash=4pt 4pt](0,0)(2,2)}
    \ThreeDput[normal=0 0 1](2,0,6){\psline[linewidth=1.2pt, linestyle=dashed, dash=4pt 4pt](0,0)(0,2)}
    \ThreeDput[normal=1 0 0](2,0,4){\psline[linewidth=1.2pt, linestyle=dashed, dash=4pt 4pt](0,0)(0,2)}
    \ThreeDput[normal=1 0 0](6,0,4){\psline[linewidth=1.2pt](0,0)(0,2)}
    \ThreeDput[normal=1 0 0](6,0,3){\psline[linewidth=1.2pt](0,3)(2,3)}
    \ThreeDput[normal=1 0 0](7,0,4){\psline[linewidth=1.2pt](0,0)(0,2)}
    \ThreeDput[normal=0 0 1](7,0,6){\psline[linewidth=1.2pt](0,0)(0,2)}
    \ThreeDput[normal=1 0 0](6,2,5){\psline[linewidth=1.2pt](0,1)(0,0)}
    \ThreeDput[normal=0 0 1](6,0,4){\psline[linewidth=1.2pt](0,0)(0,1.5)}
    \ThreeDput[normal=0 -1 0](3,0.5,7){\psline[linewidth=1.2pt](0,0)(9,0)}
    \ThreeDput[normal=0 -1 0](2,0,6){\psline[linewidth=1.2pt](0,0)(9,0)}
    \ThreeDput[normal=0 -1 0](0.8,0,4){\psline[linewidth=1.3pt,arrowsize=4pt 4]{<->}(0,0)(12.2,0)}
    \ThreeDput[normal=0 -1 0](4,0,5.1){\LARGE I}
    \ThreeDput[normal=0 -1 0](6.5,0,5.1){\LARGE II}
    \ThreeDput[normal=0 -1 0](9,0,5.1){\LARGE III}
    \ThreeDput[normal=0 -1 0](12.7,1,3.35){\LARGE $z$ }
    \put(4.25,1.76){\footnotesize $z=0$ }
    \put(5.16,1.5){\footnotesize $z=\Lz$ }
    \put(0.6,2.75){\footnotesize $z=-\Lf$ }
    \put(7.75,0.5){\footnotesize $z=\Lz + \Lf$ }
    \ThreeDput[normal=0 1 0](10,4.1,2.9){\psline[linewidth=1.2pt, arrowsize=4pt 3]{<->}(0,0)(0,2)}
    \ThreeDput[normal=0 0 1](11.3,0,4){\psline[linewidth=1.2pt, arrowsize=4pt 3]{<->}(0,0)(0,2)}
    \put(10.75,2.5){\footnotesize $\Lx$ }
    \put(10.2,0.9){\footnotesize $\Ly$ }
\end{picture}
\caption{Geometry of the cuboid containing a piecewise continuous dielectric medium.}
\label{fig:Regions}
\end{center}
\end{figure}
To facilitate the explicit construction of a basis of relevance to the Lifshitz problem attention is restricted to a particular geometry for the cavity $U$ and its boundary $(\partial U)$. In a global Cartesian co-ordinate system $(x,y,z)$, let a cuboid cavity have its six faces composed of finite regions of the rectangular planes $f_{j}=0$ ($j=1\ldots 6$) where
\begin{eqnarray*}
     \(\df f_{1}= z-(\Lz+\Lf), \; f_{2}= z+\Lf, \; f_{3}= y, \; f_{4}= y-\Ly, \; f_{5}= x, \; f_{6}=x-\Lx\)
\end{eqnarray*}
for $\Lx,\Ly,\Lz,\Lf>0$ (see figure \ref{fig:Regions}) with $\Lx\neq\Ly\neq\Lz$. The cavity is partitioned into 3 cuboid regions I, II, III by the interfaces $\Sigma_{12}$ at $z=0$ and $\Sigma_{23}$ at $z=\Lz$. Let the medium in $U$ have an anisotropic piecewise smooth homogeneous permittivity defined by
\begin{eqnarray*}
     \ep{}(z,\omega,\widehat{\underline{\xi}^{\I}},\widehat{\underline{\xi}^{\II}},\widehat{\underline{\xi}^{\III}}) \=
                        \left\{\begin{array}{lrlll}
                            \ep{0}\k^{\I}(\omega,\widehat{\underline{\xi}^{\I}}), &  \, -\Lf & \!\!\!\leq \; z \; \leq & \!\!\!0 & \;\; (\text{Region I}) \\
                            \ep{0}\k^{\II}(\omega,\widehat{\underline{\xi}^{\II}}), & \, 0 & \!\!\!\leq \; z \; \leq & \!\!\!\Lz & \;\; (\text{Region II}) \\
                            \ep{0}\k^{\III}(\omega,\widehat{\underline{\xi}^{\III}}), & \, \Lz & \!\!\!\leq \; z \; \leq & \!\!\!\Lz+\Lf & \;\; (\text{Region
                            III})
                        \end{array} \right.
\end{eqnarray*}
with real $\k^{\I},\k^{\II},\k^{\III}$ and constant real permeability $\mu=\mu_{0}$.
Given this geometry such piecewise defined functions will be written more simply as
\begin{eqnarray*}
    \epsilon &=& \sum_{\nu\in\{\reg\}} \ep{0}\k^{\nu}.
\end{eqnarray*}
The real vector parameters $\widehat{\underline{\xi}^{\nu}}$ serve to characterize the material constitutive properties of media in domain $\nu$. The condition $(\partial U)^{*}\ee_{\omega}=0$ on $\partial U$ and
\begin{eqnarray*}
\begin{tabular}{lllll}
    $\[\Sigma_{12}^{*}\hh_{\omega}\]$ & $\hspace{-0.35cm} \= 0$, & \quad &  $\[\Sigma_{12}^{*}\ee_{\omega}\]$ & $\hspace{-0.35cm} \= 0$ ,\\
    $\[\Sigma_{23}^{*}\hh_{\omega}\]$ & $\hspace{-0.35cm} \= 0$, & \quad &  $\[\Sigma_{23}^{*}\ee_{\omega}\]$ & $\hspace{-0.35cm} \= 0$ ,
\end{tabular}
\end{eqnarray*}
at the dielectric interfaces $\Sigma_{12}$ and $\Sigma_{23}$ should be compatible with the conditions to be imposed on the 1-forms
$\phi^{s,\nu}_{\omega}$ defined on $\in U_{\nu}$ used to construct 1-form solutions on $U$. Since the inhomogeneity in the permittivity $\ep{}$ is independent of the co-ordinates $x$ and $y$, the solutions for the \textit{independent 1-forms} $\pphi^{s}_{\omega,\man{N},\kk}$ with $\kk\equiv\kx,\ky$ on $U$ follow most readily by writing:
\begin{eqnarray}\label{expansion}
    \pphi^{s}_{\omega, \man{N},\kk}(\rrr) &=& \sum_{\nu\in\{\reg\}}\sum_{ \man{L}_{\nu} } \phi^{s,\nu}_{\omegaa,\kk}(x,y,z)
\end{eqnarray}
in terms of the 0-forms $\psi^{s,\nu}_{\omegaa}(z),\, \, \alpha^{s}_{\omega,\kx}(x),\,\,\beta^{s}_{\omega,\ky}(y)   $:
\begin{eqnarray}\label{CavPrepot}
    \phi^{s,\nu}_{\omegaa,\kk}(x,y,z) &=& \psi^{s,\nu}_{\omegaa}(z)\alpha^{s}_{\omega,\kx}(x)\beta^{s}_{\omega,\ky}(y)\, dz
\end{eqnarray}
where
\begin{eqnarray*}\label{cavity}
    \phi^{\TM}_{\omegaa,\kk}(x,y,z) &=& \psi^{\TM}_{\omegaa}(z)\sin(\kx x)\sin(\ky y)\, dz \\
    \phi^{\TE}_{\omegaa,\kk}(x,y,z) &=& \psi^{\TE}_{\omegaa}(z)\cos(\kx x)\cos(\ky y)\, dz
\end{eqnarray*}
and
\begin{eqnarray}\label{ode}
    \frac{d^{2}\psi^{s,\nu}_{\omegaa}}{dz^{2}} &=& \(\kx^{2} + \ky^{2} - \frac{\k^{\nu}\omega^{2}}{\cc^{2}}\)\psi^{s,\nu}_{\omegaa}
\end{eqnarray}
on $U_{\nu}$. The additional labels $\man{N}$ and $\man{L}_{\nu}$  will prove useful in the next section where the independent modes will be
synthesized from general solutions of ordinary differential equations in the regions $U_{\nu}$ with relative permittivity
$\k^{\nu}(\omega,\widehat{\underline{\xi}^{\nu}})$.
It follows from (\ref{CavPrepot}) that for any plane $\Sigma_{0}$ at $z=z_{0}$ in the cavity:
\begin{eqnarray*}
\begin{tabular}{lllll}
    $\Sigma_{0}^{*}\ee^{\TE}_{\omegaa,\kk}$ & $\hspace{-0.35cm} \= 0$ & \quad \text{if} \quad & $\psi^{\TE}_{\omegaa}(z_{0})$ & $\hspace{-0.35cm} \= 0$, \\
    $\Sigma_{0}^{*}\bb^{\TE}_{\omegaa,\kk}$ & $\hspace{-0.35cm} \= 0$ & \quad \text{if} \quad & ${\psi'}^{\TE}_{\omegaa}(z_{0})$ & $\hspace{-0.35cm} \= 0$, \\
    $\Sigma_{0}^{*}\ee^{\TM}_{\omegaa,\kk}$ & $\hspace{-0.35cm} \= 0$ & \quad \text{if} \quad & ${\psi'}^{\TM}_{\omegaa}(z_{0})$ & $\hspace{-0.35cm} \= 0$, \\
    $\Sigma_{0}^{*}\bb^{\TM}_{\omegaa,\kk}$ & $\hspace{-0.35cm} \= 0$ & \quad \text{if} \quad & $\psi^{\TM}_{\omegaa}(z_{0})$ & $\hspace{-0.35cm} \= 0$ .
\end{tabular}
\end{eqnarray*}
Thus, if the cavity walls are perfectly conducting $\psi^{s,\nu}_{\omegaa}(z), \alpha^{s}_{\omega,\kx}(x),\beta^{s}_{\omega,\ky}(y)$ must satisfy the boundary conditions
\begin{eqnarray}\label{PECCond}
\begin{tabular}{rllrl}
    $\psi^{\TE}_{\omegaa}(\Lz+\Lf)$    & $\hspace{-0.35cm} \= 0$, & \quad & $\psi^{\TE}_{\omegaa}(-\Lf)$        & $\hspace{-0.35cm} \=0$, \\
    ${\psi'}^{\TM}_{\omegaa}(\Lz+\Lf)$ & $\hspace{-0.35cm} \= 0$, & \quad & ${\psi'}^{\TM}_{\omegaa}(-\Lf)$     & $\hspace{-0.35cm} \=0$, \\
    $\alpha^{\TTM}_{\omega,\kx}(0) $   & $\hspace{-0.35cm} \= 0$, & \quad & $\alpha^{\TTM}_{\omega,\kx}(\Lx)$   & $\hspace{-0.35cm} \=0$, \\
    ${\alpha'}^{\TE}_{\omega,\kx}(0)$  & $\hspace{-0.35cm} \= 0$, & \quad & ${\alpha'}^{\TE}_{\omega,\kx}(\Lx)$ & $\hspace{-0.35cm} \=0$, \\
    $\beta^{\TTM}_{\omega,\ky}(0)$     & $\hspace{-0.35cm} \= 0$, & \quad & $\beta^{\TTM}_{\omega,\ky}(\Ly)$    & $\hspace{-0.35cm} \=0$, \\
    ${\beta'}^{\TE}_{\omega,\ky}(0)$   & $\hspace{-0.35cm} \= 0$, & \quad & ${\beta'}^{\TE}_{\omega,\ky}(\Ly)$  & $\hspace{-0.35cm} \=0$,
\end{tabular}
\end{eqnarray}
where prime denotes the derivative of a function with respect to its argument and furthermore, at each perfectly insulating interface $\Sigma=\Sigma_{12},\Sigma_{23}$, if
\begin{eqnarray}\label{IFCond}
\begin{tabular}{lllll}
    $\displaystyle \left. \df \[\psi^{\TE}_{\omegaa}\]\right|_{\Sigma}$    & $\hspace{-0.35cm} \= 0$ & \quad \text{then} \quad & $\displaystyle\[\Sigma^{*}\ee^{\TE}_{\omegaa,\kk}\]$ & $\hspace{-0.35cm} \= 0$, \\
    $\displaystyle\left. \df \[{\psi'}^{\TE}_{\omegaa}\]\right|_{\Sigma}$ & $\hspace{-0.35cm} \= 0$ & \quad \text{then} \quad & $\displaystyle\[\Sigma^{*}\hh^{\TE}_{\omegaa,\kk}\]$ & $\hspace{-0.35cm} \= 0$, \\
    $\displaystyle\left. \df \[{\psi'}^{\TM}_{\omegaa}\]\right|_{\Sigma}$ & $\hspace{-0.35cm} \= 0$ & \quad \text{then} \quad & $\displaystyle\[\Sigma^{*}\ee^{\TM}_{\omegaa,\kk}\]$ & $\hspace{-0.35cm} \= 0$, \\
    $\displaystyle\left. \df \[\psi''^{\TM}_{\omegaa}\]\right|_{\Sigma}$  & $\hspace{-0.35cm} \= 0$ & \quad \text{then} \quad & $\displaystyle \[\Sigma^{*}\hh^{\TM}_{\omegaa,\kk}\]$ & $\hspace{-0.35cm} \= 0$,
\end{tabular}
\end{eqnarray}
since the permeability is independent of position $U_{\nu}$. Then (\ref{HelmMaxDom}) is satisfied with\footnote{When $\nx$ and $\ny$ are simultaneously zero, all modes vanish. When $\nx$ or $\ny$ are zero, then all TM modes vanish.}
\begin{eqnarray}\label{kcnds}
    \kx &=& \frac{\nx\pi}{\Lx} \qquadand \ky \= \frac{\ny\pi}{\Ly} \qquad \nx,\ny\in\{0,1,2,\ldots\}.
\end{eqnarray}
With
\begin{eqnarray*}
    k^{2} &\equiv& \kx^{2} + \ky^{2} \= \frac{\nx^{2}\pi^{2}}{\Lx^{2}} + \frac{\ny^{2}\pi^{2}}{\Ly^{2}},
\end{eqnarray*}
introduce real constant positive dimensionless parameters $R,\,\Omega,\,\underline{\lambda}=(\lambda_{1},\, \lambda_{2}, \, \lambda)$ by the relations
\begin{eqnarray}\label{NDrom}
    \Lx &=& \lambda_{1}\Lz, \quad \Ly \= \frac{\lambda_{1}}{\lambda_{2}}\Lz, \quad  \Lf \= \lambda\Lz, \quad \frac{\omega}{\cc} \= \frac{\Omega}{\Lz}, \quad k \= \frac{R}{\Lz},
\end{eqnarray}
so
\begin{eqnarray}\label{rRdef}
     R^{2} &=& \frac{\pi^{2}}{\lambda_{1}^{2}}\(\nx^{2} + \lambda_{2}^2 \ny^{2}\) \qquadand \man{K}^{\nu}(\Omega, \underline{\xi}^{\nu}) \= \k^{\nu}(\omega,\widehat{\underline{\xi}^{\nu}}).
\end{eqnarray}
The parameters $\underline{\xi}^{\nu}$ are non-dimensionalized parameters constructed from the $\widehat{\underline{\xi}^{\nu}}$.
Then (\ref{ode}) becomes, with $\hpsi^{s,\nu}_{\Omega,\man{L}_{\nu}}(Z) =  \psi^{s,\nu}_{\omega,\man{L}_{\nu}}(z)$:
\begin{eqnarray}\label{fgDE2}
    \SOdiff{ \hpsi^{s,\nu}_{\Omega,\man{L}_{\nu}}(Z)}{Z} &=& {\(R^{2}-\kr^{\nu}\Omega^{2}\)} \,  \hpsi^{s,\nu}_{\Omega,\man{L}_{\nu}}(Z)
\end{eqnarray}
where $Z=z/\Lz$. Thus $\Sigma_{12}$ is at $Z=0$ and $\Sigma_{23}$ is at $Z=1$.
The nature of the general solution to (\ref{fgDE2}) depends on whether:
\begin{eqnarray*}
\begin{tabular}{rll}
    (i)   & \; $R^{2} - \kr^{\nu}\Omega^{2}$ & $\hspace{-0.35cm} > 0$ \\
    (ii)  & \; $R^{2} - \kr^{\nu}\Omega^{2}$ & $\hspace{-0.35cm} < 0$ \\
    (iii) & \; $R^{2} - \kr^{\nu}\Omega^{2}$ & $\hspace{-0.35cm} \= 0$.
\end{tabular}
\end{eqnarray*}
Introducing further the real expressions{\footnote{In the following it will become clear that $\Omega$ and hence $\chinu$ and $\zetanu$ depend on $s\in\{\text{TE,TM}\}$. To economize on notation this implicit dependence will not always be made explicit in the following.}}
\begin{eqnarray}\label{zetachiDEF}
    \zetanu \equiv \sqrt{R^{2} - \kr^{\nu}\Omega^{2}} \;> 0 \qquadand \chinu \equiv \sqrt{\kr^{\nu}\Omega^{2} - R^{2}} \;> 0
\end{eqnarray}
for $\nu=$ I, II, III, one has
\begin{eqnarray*}\renewcommand{\arraystretch}{2.4}
\begin{tabular}{llll}\
    $\displaystyle \SOdiff{\MMD{\hpsi}{s}{\nu}{\Omega}{E}}{Z}$ & $\hspace{-0.35cm} \= \zetanu^2 \,\MMD{\hpsi}{s}{\nu}{\Omega}{E}$ & \text{for} \quad & $R^2>\kr^{\nu} \,\Omega^2$, \\
    $\displaystyle \SOdiff{\MMD{\hpsi}{s}{\nu}{\Omega}{O}}{Z}$ & $\hspace{-0.35cm} \= -\chinu^2 \,\MMD{\hpsi}{s}{\nu}{\Omega}{O}$ & \text{for} \quad & $R^2<\kr^{\nu} \,\Omega^2$, \\
    $\displaystyle \SOdiff{\MMD{\hpsi}{s}{\nu}{\Omega}{L}}{Z}$ & $\hspace{-0.35cm} \= 0$ & \text{for}\quad & $R^2\=\kr^{\nu}\,\Omega^2$,
\end{tabular}
\end{eqnarray*}\renewcommand{\arraystretch}{1.7}
with general solutions:
\begin{eqnarray}\label{HelmGenSol}
\begin{tabular}{llll}
    $\MMD{\hpsi}{s}{\nu}{\Omega}{E}{(Z)}$ & $\hspace{-0.35cm} \= \MMD{P}{s}{\nu}{\Omega}{E}\, \cosh(\zetanu Z)$ & $+$ & $\MMD{Q}{s}{\nu}{\Omega}{E}\, \sinh(\zetanu Z)$,\\
    $\MMD{\hpsi}{s}{\nu}{\Omega}{O}{(Z)}$ & $\hspace{-0.35cm} \= \MMD{P}{s}{\nu}{\Omega}{O}\, \cos(\chinu Z)$ & $+$ & $\MMD{Q}{s}{\nu}{\Omega}{O}\, \sin(\chinu Z)$,\\
    $\MMD{\hpsi}{s}{\nu}{\Omega}{L}{(Z)}$ & $\hspace{-0.35cm} \= \MMD{P}{s}{\nu}{\Omega}{L}\, $ & $+$ & $\MMD{Q}{s}{\nu}{\Omega}{L}\,Z $,
\end{tabular}
\end{eqnarray}
respectively in terms of complex constants $\MD{P}{s}{\nu}{\Omega}{L}, \, \MD{Q}{s}{\nu}{\Omega}{L}\!$ with $\nu\in\{\text{I, II, III}\}$ and labels ${ \cal L}\in \{ \text{E,O,L}  \}$ designating a hyperbolic, oscillatory and affine-linear $Z$ behaviour of $ \psi^{s,\nu}_{\Omega, \man{L}_{\nu}}$ on $U_{\nu}$. Thus:
\begin{eqnarray}\label{chizetadef}
\begin{tabular}{rllll}
    In region I:   & \hspace{1cm} $\chiI$   & $\hspace{-0.35cm} \= \sqrt{\kr^{\I} \Omega^2 - R^2}$,   \quad & $\zetaI$   & $\hspace{-0.35cm} \= \sqrt{R^2 -\kr^{\I} \Omega^2 } $   \\
    In region II:  & \hspace{1cm} $\chiII$  & $\hspace{-0.35cm} \= \sqrt{\kr^{\II} \Omega^2 - R^2}$,  \quad & $\zetaII$  & $\hspace{-0.35cm} \= \sqrt{R^2 - \kr^{\II}\Omega^2 }$    \\
    In region III: & \hspace{1cm} $\chiIII$ & $\hspace{-0.35cm} \= \sqrt{\kr^{\III} \Omega^2 - R^2}$, \quad & $\zetaIII$ & $\hspace{-0.35cm} \= \sqrt{R^2 - \kr^{\III} \Omega^2}$.
\end{tabular}
\end{eqnarray}
The piecewise defined functions $\psi^{s}_{\Omega,\man{N}}$ on $U$ must be synthesized by matching suitable combinations of the
$\psi^{s,\nu}_{\Omega, \man{L}_{\nu}}$ across the interfaces. There are $3^3=27$ possible combinations corresponding to the 27 possible ways of making a 3-letter word ${\cal N}$ out of the letters in $ \{ \text{E,O,L} \} $. However not all 3-letter configurations are, in general, compatible with any particular choice of real $\kr^{\nu}$. For example, for a configuration with $\kr^{\I}(\Omega,\underline{\xi})=\kr^{\III}(\Omega,\underline{\xi})\equiv \kr(\Omega,\underline{\xi})$, one cannot have
both $\zetanu>0$ and $\chinu>0$, or $\zetanu>0$ and $\chinu=0$, or $\chinu>0$ and $\zetanu=0$. That is, only modes of the same type are permitted in the dielectric regions I and III. This reduces the number of configurations to 9. Furthermore, with $\kr^{\nu} \neq 1$  the configuration $\man{N}=(\text{LLL} )$ is not possible since the  affine-linear solutions defined by $\chinu=0$ with $\Omega=R/ \sqrt{\kr^{\nu}}$ are incompatible with $\chiII=0$ with $\Omega=R$.  The remaining possible allowed $\man{N}$ configurations belong to the set:
\begin{eqnarray*}
     \{ \text{OOO, OEO, OLO, EOE, EEE, ELE, LOL, LEL} \}.
\end{eqnarray*}
In this notation the 1-forms $\pphi^{s}_{\omega,{\cal N },\kk}(x,y,z)$ may be said to belong to the spectral family $[{\cal N}
]_z\,[\text{O}]_x\,[\text{O}]_y  $. As a result of this simplification, it proves expedient to motivate the general discussion by explicit reference  to configurations having $\kr^{\I}(\Omega,\xi)=\kr^{\III}(\Omega,\xi)\equiv \kr(\Omega,\xi)$, $\kr^{\II}=1$ with $\kr$ dependent upon a real single dimensionless material constant $\xi$, parameterising the medium and:
\begin{eqnarray*}
\begin{tabular}{lll}
    $\kr$ & $\hspace{-0.35cm} \= \xi$ & \qquad (describing a constant dielectric model) \\
    $\kr$ & $\displaystyle \hspace{-0.35cm} \= 1 - \frac{\xi}{\Omega^{2}}$ & \qquad (describing a metallic ``plasma'' model).
\end{tabular}
\end{eqnarray*}

\section{Spectra Associated with Allowed Configurations in the Cavity }\label{Sect:SpectraCavity}
The discrete cavity spectra $\{\Omega\}$ associated with the spectral modes constructed from the forms
$\MMD{\pphi}{s}{}{\omega,\man{N},\kk}{}(x,y,z)$ are determined by the boundary and interface conditions (\ref{PECCond}) and (\ref{IFCond}). For each $s\in\{\text{TE,TM}\}$, these conditions give rise to a homogeneous, algebraic system for the constants
$\MMD{P}{s}{\nu}{\Omega}{\man{L}},\MMD{Q}{s}{\nu}{\Omega}{\man{L}}$ in (\ref{HelmGenSol}) that determine the field spectral modes up to an overall scaling. Such constants are non-trivial provided the determinant of each homogeneous system is zero:
\begin{eqnarray}\label{DetCond}
    \man{D}^{s}_{\man{N},\mbox{\tiny $\nx,\ny$}}( \Omega, \underline{\lambda}, \xi ) &=& 0.
\end{eqnarray}
For fixed $\underline{\lambda}, \xi$, each real root of this (in general) transcendental equation is an element of a discrete frequency spectrum associated with the configuration $(s,\man{N},\nx,\ny)$. For any strictly non-vanishing
factor $\gamma$, the expression $ \gamma\, \man{D}^{s}_{\man{N},\mbox{\tiny $\nx,\ny$}}( \Omega, \underline{\lambda},\xi )  $ will be referred to as an {\it $s$-spectrum generator} for such a configuration and  (\ref{DetCond}) as an {\it implicit global dispersion relation} for modes in the cavity. The homogeneous algebraic systems and associated cavity discrete spectrum generators are summarized in \ref{BoundAppendix}, where it is understood that the real expressions $\chi_{\nu},\zeta_{\nu}$ are given in terms of $\Omega,R,\xi$ by the definitions in the previous section.

\subsection{Completeness for the  Discrete Cavity Modes}\label{Sect:CavityCompleteness}
\def\YY{ {\cal Y}}
\def\bfM{{ \underline{M} }}
\def\JJ{{j_{{\cal N}\kk}  }}

Given the results of \ref{BoundAppendix}, the cuboid geometry and a specific piecewise defined permittivity, it is possible, in principle, to enumerate a complete set of orthogonal modes with respect to the inner product (\ref{IPROD}). Let the independent modes be labelled by $\bfM = [ {\cal N},j,n_x,n_y]$ where the positive integer
$j$ enumerates {\it all solutions} (including possible degeneracies)  of the homogeneous boundary condition system for each orthogonal spectral family labelled by $ {\cal N}$. For each $\bfM$ the discrete spectral parameters $ \Omega^{s}_{\JJ}$ satisfy (\ref{DetCond}). A set of basis 1-forms $\YY_{\tm{\bfM}}$ in the span of the 1-form modes $ \pphi^s_{\omega, {\cal N }, \kk} $ constructed to satisfy
\begin{eqnarray*}
    (\YY_{\tm{\bfM}},\YY_{\tm{\bfM}'})_{\tm{U}} &=& \delta_{\tm{\bfM}\,\tm{\bfM}'}\,w_{\tm{\bfM}}
\end{eqnarray*}
in terms of the Kronecker $\delta$ and a set of arbitrary positive normalization constants $\{w_{\tm{\bfM}}\}$ is said to be orthogonal.
Define the operators
\begin{eqnarray*}
\begin{tabular}{llrll}
    $\wt{\pphi}_{\tm{\bfM}}: \Lambda^1(U) \to \Lambda^1(U)$ & \quad and \quad & $\Delta_{(x,y,z)}: \Lambda^1(U)$ & $\!\!\to\!\!$ & $\Lambda^1(U)$ \\
    & & $\theta$ & $\!\!\mapsto\!\!$ & $\Delta_{(x,y,z)}[\theta]\;=\;\theta(x,y,z)$
\end{tabular}
\end{eqnarray*}
mapping 1-forms on $U$ to 1-forms on $U$ with $\wt{\pphi}_{\tm{\bfM}}\equiv (\pphi_\tm{\bfM}, -)_{\tm{U}}$. Abbreviating the mode summation operator with
\begin{eqnarray*}
    \sum_{\tm{\bfM}}\equiv \sum_s \sum_{\cal N }\sum_{\nx} \sum_{\ny} \sum_j
\end{eqnarray*}
the discrete cavity 1-form modes are said to be complete with this mode normalization when
\begin{eqnarray}\label{BoundedCR}
    \sum_{\tm{\bfM}} \frac{\YY_{\tm{\bfM}}(x,y,z) } {w_{\tm{\bfM}} }\,\, \wt{\YY}_{\tm{\bfM}} &=& \Delta_{(x,y,z)}.
\end{eqnarray}
For each $s$ and $\man{N}$, the roots $\Omega^{s}_{\JJ}$ of (\ref{DetCond}) for fixed cuboid geometry defined by $\underline{\lambda}$ and fixed medium permittivity in regions I and III defined by $\xi$ cannot be determined analytically. Numerically, they can be determined to arbitrary accuracy by locating the zeroes of the determinant in (\ref{DetCond}) in a three-dimensional phase space with co-ordinates $(N_{x},N_{y},\Omega)$. For fixed $s,\man{N},\underline{\lambda},\xi$, (\ref{DetCond}) describes a family of two-dimensional \textit{spectral surfaces} in this phase space and the roots $\Omega^{s}_{\JJ}$ occur where the vertical line $N_{x}=n_{x} \in \mathbb{Z}^{+},\,N_{y}=n_{y} \in \mathbb{Z}^{+}$ intersects this family of surfaces. For the symmetric configuration with $\Lx=\Ly$, all states with the same value of $\nx^{2} + \ny^{2}$ (i.e. the same $k^{2}=\kx^{2}+\ky^{2}$) have the same discrete set of roots. The structure of the spectral surfaces is such that in general the value of the roots increases with $k$ for fixed $s,\man{N},\underline{\lambda},\xi$.

The description of the \textit{quantum} states of the electromagnetic field in the cavity with a medium is based on a choice of Hamiltonian and a Hilbert space for the system. The natural choice of these is motivated by the well established quantization of the electromagnetic field in free space in terms of a collection of harmonic oscillators and a choice of ground state. If each quantum mode of the field is associated with the ground state of a single harmonic oscillator of any angular frequency $\omega$, it is assigned an energy $\frac{1}{2}\hbar \omega$. In cavity QED, this procedure is extended to describe modes subject to confining boundary conditions and the Hamiltonian is modified to accommodate the constitutive properties of the confined medium. Thus, the lowest energy associated with each allowed discrete mode of any angular frequency $\omega_{\tm{\bfM}}$ becomes $\frac{1}{2}\hbar \omega_{\tm{\bfM}}$. Since the number of modes is infinite and, in general, $\omega_{\tm{\bfM}}$ grows with $k$, no meaning can be attached to the lowest energy associated with \textit{all} modes. However, meaning can be attached to the lowest energy difference associated with all modes in cavities with the {\it same volume} but different media contents. To compute this difference, one requires a complete knowledge of the spectra of the two systems. In situations where two spectra are smoothly connected in the space of all spectra, one can in principle compute the lowest energy difference. For the cuboid geometry under discussion, if $\Lx,\Ly,\Lz,\Lf$ are kept fixed, $\{\omega_{\tm{\bfM}}(\Lx,\Ly,\Lz,\Lf, \xi)\}$ denote a set of curves in the space of spectra. If these curves smoothly connect a material configuration described by $\xi$ to one described by $\xi^{\tt{ref}}$, no new modes arise in the continuation. Thus, the ground state energy of a cuboid containing a medium with permittivity characterized by $\xi$ relative to one containing a medium with permittivity characterized by $\xi^{\tt{ref}}$ is finite and may be represented by the sum
\begin{eqnarray*}
    \mbox{\small $\displaystyle \man{E}(\Lx,\Ly,\Lz,\Lf,\xi,\xi^{\tt{ref}} ) \= \frac{\hbar}{2} \sum_{\tm{\bfM}} \[ \df \omega_{\tm{\bfM}}(\Lx,\Ly,\Lz,\Lf,\xi) - \omega_{\tm{\bfM}}(\Lx,\Ly,\Lz,\Lf,\xi^{\tt{ref}}) \]$}
\end{eqnarray*}
if convergent. In particular, one may consider the energy difference relative to a finite volume empty cuboid. For piecewise homogeneous media, such an energy difference gives rise to a stress field in the medium and a normal pressure difference across the interfaces in the medium of value
\begin{eqnarray}\label{pressureBoundDef}
    \man{P}(\Lx,\Ly,\Lz,\Lf,\xi,\xi^{\tt{ref}}) &=& -\pdiff{}{\Lz}\,\frac{\man{E}(\Lx,\Ly,\Lz,\Lf,\xi,\xi^{\tt{ref}} )}{\Lx\Ly}.
\end{eqnarray}
Such a stress field is transmitted to a normal pressure on the end faces of the cuboid. As the volume of the cuboid increases, the number of states less than any fixed angular frequency, in general, increases and any attempt to compare with configurations that approach open systems
inevitably requires a knowledge of the density of states of such systems. In principle, this can be deduced from the behaviour of
$\omega_{\tm{\bfM}}(\Lx,\Ly,\Lz,\Lf,\xi)$ as the geometrical parameters $\Lx,\Ly,\Lf$ increase for fixed $\xi$ and is the subject of extensive
mathematical analysis following Weyl's pioneering work on the spectra of the wave equation in open domains \cite{baltes1976spectra}. In practice, various regularization techniques have been devised for both closed and open configurations in order to isolate various divergences arising from non-convergent mode summations. An alternative approach to open systems, pursued in the next section, is to explore the structure of the \textit{continuous} mode spectra for the harmonic electromagnetic fields ab initio in  $\real^{3}$ guided by the methodology used to determine the structure of the discrete mode spectra discussed in this section for the confined modes in the cuboid cavity.

\section{Global and Local Dispersion Relations Associated with Allowed Configurations in a Stratified Open Medium}\label{Sect:DispOpen}
In a stratified \textit{open} (unconfined) medium one may classify distinct modes by the behaviour of distinct families of electromagnetic fields in the medium. In many cases, harmonic families describe modes with continuous real  frequencies. There are no physical (confining) boundaries hence no physical boundary conditions to impose. Since the medium is homogeneous in $x$ and $y$ the behaviour of harmonic fields in these variables is oscillatory as in the  (closed) cavity but now the wave numbers $\kx$ and $\ky$ can be arbitrary non-zero real numbers. Thus in the open situation the dimensionless discrete variable in (\ref{rRdef})  is replaced by the dimensionless continuous variable $R=k L_z$ where $k^2=k_x^2+k_y^2$. As before the scale factor $L_z$ can be used to define the continuous dimensionless angular frequency $\Omega= \omega\Lz / c$ and the dimensionless coordinate $Z=z/L_z$.

The behaviour of the modes as a function of $Z$ is determined by the same interface conditions as in the cavity case plus the condition that all fields are finite as $|Z|\rightarrow\infty$. A natural classification of the modes is to situations where they are oscillatory (propagating) in $Z$ in regions I and III, viz:
\begin{eqnarray*}
    \mbox{Category P: } \qquad [\text{OOO}]_z[\text{O}]_x[\text{O}]_y,\quad [\text{OEO}]_z[\text{O}]_x[\text{O}]_y,\quad
    [\text{OLO}]_z[\text{O}]_x[\text{O}]_y
\end{eqnarray*}
and to situations where the modes have damped exponential (evanescent) behaviour for large $\vert Z \vert $ in regions I and III,  viz:
\begin{eqnarray*}
    \mbox{Category E: } \qquad [\text{EOE}]_z[\text{O}]_x[\text{O}]_y,\quad [\text{EEE}]_z[\text{O}]_x[\text{O}]_y,\quad
    [\text{ELE}]_z[\text{O}]_x[\text{O}]_y.
\end{eqnarray*}
It will transpire that not all these modes can exist when one takes a symmetric configuration with $\k(\omega,\wh{\xi})>0$.
For category P modes one may choose any non-zero real values of $\omega,\kx,\ky$ that, for a given $\k^{\nu}$, are consistent with $\chinu$ and $\zetanu$ being positive for $\nu=$ I, II, III. The resulting relations between $\omega,\kx,\ky,\kz$ where $\kz=\chinu/\Lz$ or $\kz=\zetanu/\Lz$ may be termed {\it local dispersion relations}. For example for the category P modes $[\text{OOO}]_z[\text{O}]_x[\text{O}]_y$ for all $\omega,\kx,\ky$ that yield real positive $\chinu$, one has the local relations:
\begin{eqnarray*}
\begin{tabular}{lll}
    $\displaystyle \frac{\k\omega^2}{\cc^2}$ & $\displaystyle \hspace{-0.35cm} \= \kx^2+\ky^2 + \frac{\chi_{\I}^{2}}{\Lz^{2}}$   & \quad in I, \\
    && \\
    $\displaystyle \frac{\omega^2}{\cc^2}$   & $\displaystyle \hspace{-0.35cm} \= \kx^2+\ky^2 + \frac{\chi_{\II}^{2}}{\Lz^{2}}$  & \quad in II, \\
    && \\
    $\displaystyle \frac{\k\omega^2}{\cc^2}$ & $\displaystyle \hspace{-0.35cm} \= \kx^2+\ky^2 + \frac{\chi_{\III}^{2}}{\Lz^{2}}$ & \quad in III.
\end{tabular}
\end{eqnarray*}
Modes satisfying these conditions with constant $\k=\xi$ describe harmonic plane fronted waves propagating in directions $(\kx,\ky,\chinu/\Lz)$ with speed $\cc/\sqrt{\k}$ in regions $\nu=$ I, III and in directions $(\kx,\ky,\chi_{\II}/\Lz)$ with speed $\cc$ in region II.
Modes in category E are characterized by arbitrary non-zero real numbers $\kx,\ky$ but are exponentially damped as $\vert Z\vert \to \infty$. This energy bound condition may imply that only certain real $\omega$ are compatible with the interface conditions. However since the $\kx,\ky$ are continuous variables the compatible frequencies $\omega^s_{\cal N}= \wt{\omega}^s_{\cal N}(\kx,\ky)$ constitute a family of continuous spectra. Such relations may be termed {\it global dispersions relations} since they make no reference to any particular region in the medium. They describe harmonic waves with propagating characteristics in the $x$ and $y$ directions but are exponentially damped as $\vert Z \vert$ becomes large. Such modes are sometimes referred to as {\it dielectric waveguide} modes.

For category P modes it is convenient to introduce the complex solutions:
\begin{eqnarray*}
    \MMD{\hpsi}{s}{\nu}{\Omega}{\O}{(Z)} &=& \MMD{P}{s}{\nu}{\Omega}{\O}\, e^{-i\chi_{\nu} Z} + \MMD{Q}{s}{\nu}{\Omega}{\O}\, e^{i\chi_{\nu} Z}
\end{eqnarray*}
with the form of $\MMD{\hpsi}{s}{\II}{\Omega}{\man{L}_{\II}}{(Z)}$ determined by the label ${\cal L} $ in region II. For each $s \in
\{\text{TE,TM}\}$ the four interface conditions imply four linear relations among the six amplitudes in
\begin{eqnarray*}
    \man{C}^{s} &=& \left\{\df \MMD{P}{s}{\I}{\Omega}{\O}, \MMD{Q}{s}{\I}{\Omega}{\O}, \MMD{P}{s}{\II}{\Omega}{\man{L}_{\II}},
    \MMD{Q}{s}{\II}{\Omega}{\man{L}_{\II}}, \MMD{P}{s}{\III}{\Omega}{\O}, \MMD{Q}{s}{\III}{\Omega}{\O}  \right\}.
\end{eqnarray*}
and for each $s$, any four elements of $\man{C}^{s}$ can be expressed in terms of the remaining two elements of $\man{C}^{s}$. This two-fold
degeneracy can be expressed in terms of a basis of solutions, representing  scattering of a plane wave incident from the left or the right. Such states are defined from a general solution of the linear relations in terms of $\MMD{Q}{s}{\I}{\Omega}{\O},\MMD{P}{s}{\III}{\Omega}{\O}$ by employing the particular substitutions:
\begin{eqnarray}\label{P_RL}
\begin{tabular}{lllllllrlrll}
    $\man{C}^{s}_{\tt{R}}$ & $\equiv$ & $\displaystyle \left\{\df \MMD{P}{s}{\I}{\Omega}{\O}\right.$ & $\!\!\!=\!\!\!$ & $ R^{s,\tt{R}}_{\man{N}}$, & $\MMD{Q}{s}{\I}{\Omega}{\O}{}$ & $\!\!\!=\!\!\!$ & $\; 1, \;\; \MMD{P}{s}{\III}{\Omega}{\O}{}$ & $\!\!\!=\!\!\!$& $\; 0, \;\; \MMD{Q}{s}{\III}{\Omega}{\O}{}$ & $\!\!\!=\!\!\!\!$ & $\displaystyle \left. \df T^{s,\tt{R}}_{\man{N}} \right \}$ \\
    $\man{C}^{s}_{\tt{L}}$ & $\equiv$ & $\displaystyle \left\{\df \MMD{P}{s}{\I}{\Omega}{\O}{} \right.$ & $\!\!\!=\!\!\!$ & $ T^{s,\tt{L}}_{\man{N}}$, & $\MMD{Q}{s}{\I}{\Omega}{\O}{}$ & $\!\!\!=\!\!\!$ & $0, \;\; \MMD{P}{s}{\III}{\Omega}{\O}{}$ & $\!\!\!=\!\!\!$ & $1, \;\; \MMD{Q}{s}{\III}{\Omega}{\O}{}$ & $\!\!\!=\!\!\!\!$ & $\displaystyle \left. \df R^{s,\tt{L}}_{\man{N}}  \right\}$.
\end{tabular}
\end{eqnarray}
The complex pairs $(R^{s,\tt{R}}_{\man{N}}, R^{s,\tt{L}}_{\man{N}}), \, (T^{s,\tt{R}}_{\man{N}},T^{s,\tt{L}}_{\man{N}})$ denote complex reflection and transmission coefficients where R,L designate configurations describing waves propagating from region I to III and from region III to I respectively. The category E modes are characterized by modes that are exponentially damped as functions of $Z$ as $\vert Z \vert \rightarrow \infty$:
\begin{eqnarray}\label{PsiUnboundB}
\begin{tabular}{rl}
    $\MMD{\hpsi}{s}{\I}{\Omega}{\E}{(Z)}$ & $\hspace{-0.3cm} \= \MMD{Q}{s}{\I}{\Omega}{\E}\, e^{\zetaI Z}$ \\
    $\MMD{\hpsi}{s}{\III}{\Omega}{\E}{(Z)}$ & $\hspace{-0.3cm} \= \MMD{P}{s}{\III}{\Omega}{\E}\, e^{-\zetaIII Z}$
\end{tabular}
\end{eqnarray}
with the form of $\MMD{\hpsi}{s}{\II}{\Omega}{\man{L}_{\II}}{(Z)}$  determined by the label $ {\cal L} $ in region II.
A cursory examination of the structure of the complex coefficients describing the category P and category E mode solutions for the open domain
(given in \ref{OpenAppendix}) show them to be correlated by symbolic transformations that map $\pm i\chiI \leftrightarrow \zetaI$ and $\pm i\chiII\leftrightarrow \zetaII$. This suggests that the \textit{real} functions of the \textit{real} variable $\Omega'$ may be values of one or more complex functions of an extension of $\Omega'$ to some complex variable $\Omega=\Omega'+i\Omega''$ and that the above transformations arise as a result of a choice of various square roots in such functions. To implement this idea, one needs to define them by suitable analytic continuations. Thus, suppose one has a real or complex function $\f(\Omega')$ of a \textit{real} variable $\Omega'$ that has a natural extension to a function $\f(\Omega)$ of a complex variable $\Omega=\Omega'+i\Omega''$. In the complex $\Omega$-plane, such a function will, in general, have branch points, zeroes and poles. If $\f(\Omega)$ is analytic (free of all singularities) on a domain $\man{S}$ of the upper half $\Omega$-plane that includes any segment $\Upsilon$ of the real $\Omega$-axis and is real on $\Upsilon$, then there exists an analytic function $\wt{\f}(\Omega)$ on $\man{S}\cup\conj{\man{S}}$ where $\conj{\man{S}}$ is the reflection of $\man{S}$ in $\Upsilon$. This analytic continuation of $\f(\Omega)$ to $\wt{\f}(\Omega)$ on $\man{S}\cup\conj{\man{S}}$ is defined by
\begin{eqnarray}\label{AContinDef}
    \wt{\f}(\Omega) &=& \left\{ \begin{array}{ll}
                                \f(\Omega), & \Omega \in \man{S} \\
                                \conj{\f}(\conj{\Omega}), & \Omega \in \conj{\man{S}}
                            \end{array}\right.
\end{eqnarray}
since it is easy to show that $\wt{\f}(\Omega)$ satisfies the Cauchy-Riemann equations on $\man{S}\cup\conj{\man{S}}$. An immediate consequence is the identity
\begin{eqnarray*}
    \oint_{C} \wt{\f}(\Omega) \, d\Omega &=& 0 \qquad \text{for all closed contours $C$ in } \man{S}\cup\conj{\man{S}}.
\end{eqnarray*}
\begin{figure}[h!]
    \centering
    \includegraphics[width=0.8\textwidth]{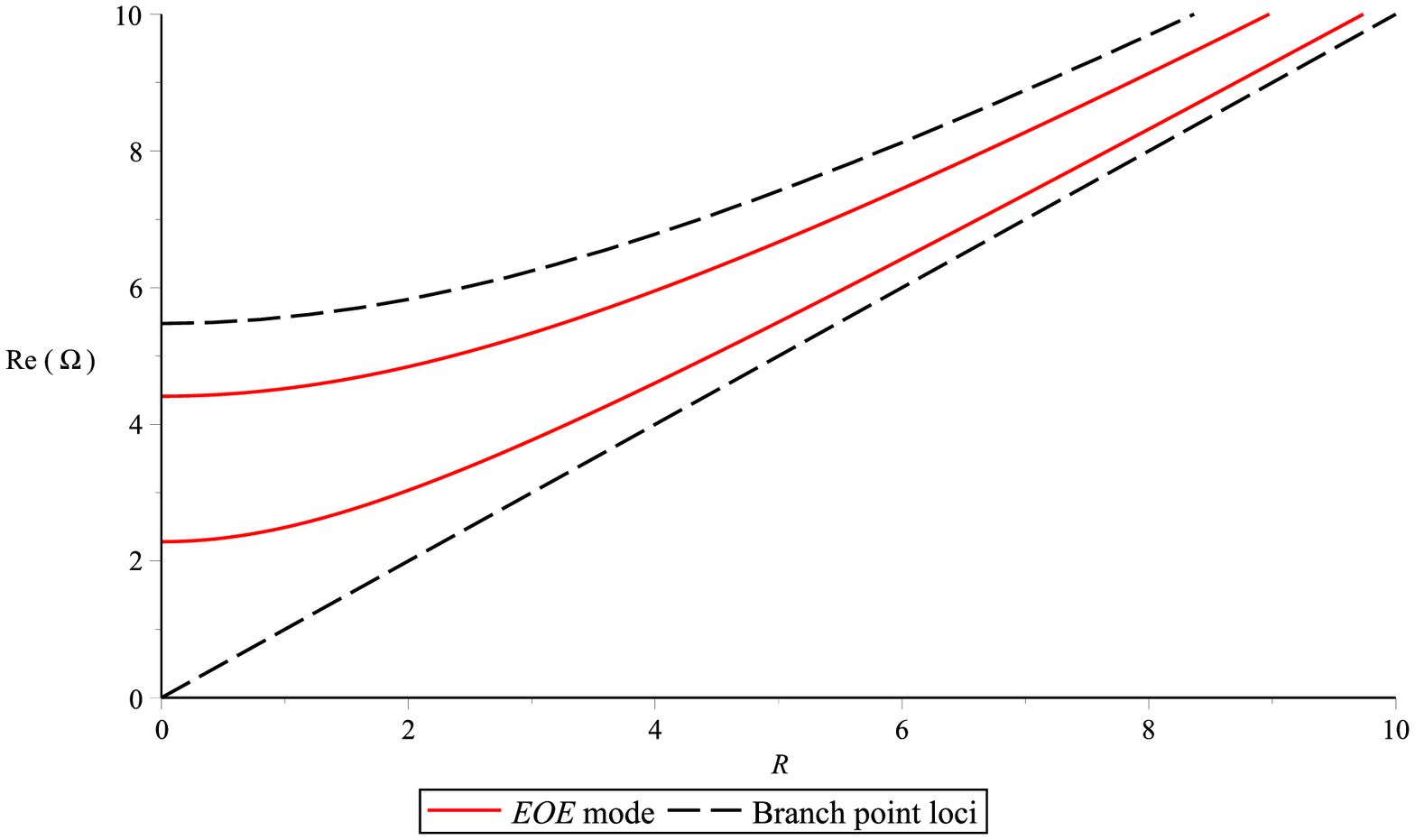}
    \caption{Location of category E poles and branch points with $\Omega'>0$ on real $\Omega$ axis for TE plasma model with $\xi=30$ are located where a vertical line at fixed $R$ intersects any red loci and dotted loci respectively. There are no EEE or ELE modes with real frequency in this model. The indicated EOE mode loci follow from ${D}^{\TTE}_{\E\O\E,\kx,\ky}( \Omega',\xi)=0$ (\ref{DTE_EOE}). }
    \label{fig:TE_PLASMA}
\end{figure}
\begin{figure}[h!]
    \centering
    \includegraphics[width=0.8\textwidth]{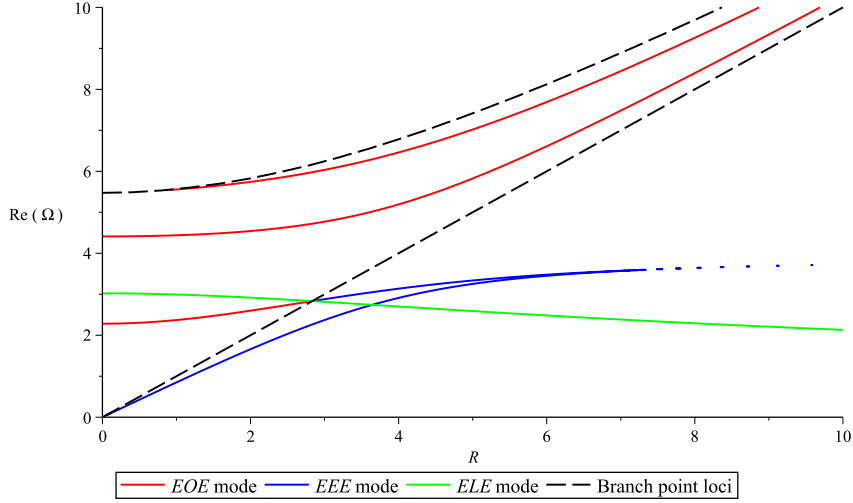}
    \caption{Location of category E poles and branch points with $\Omega'>0$ on real $\Omega$ axis for TM plasma model with $\xi=30$ are located where a vertical line at fixed $R$ intersects any red or blue loci and dashed loci respectively. The diagram exhibits one ELE mode where the green locus intersects the dashed line. The indicated EOE, EEE and ELE mode loci follow from ${D}^{\TTM}_{\E\O\E,\kx,\ky}( \Omega',\xi)=0$ (\ref{DTM_EOE}), ${D}^{\TTM}_{\E\E\E,\kx,\ky}( \Omega',\xi)=0$ (\ref{DTM_EEE}) and ${D}^{\TTM}_{\E\L\E,\kx,\ky}( \Omega',\xi)=0$ (\ref{DTM_ELE}) respectively. }
    \label{fig:TM_PLASMA}
\end{figure}
\begin{figure}[h!]
    \centering
    \includegraphics[width=0.8\textwidth]{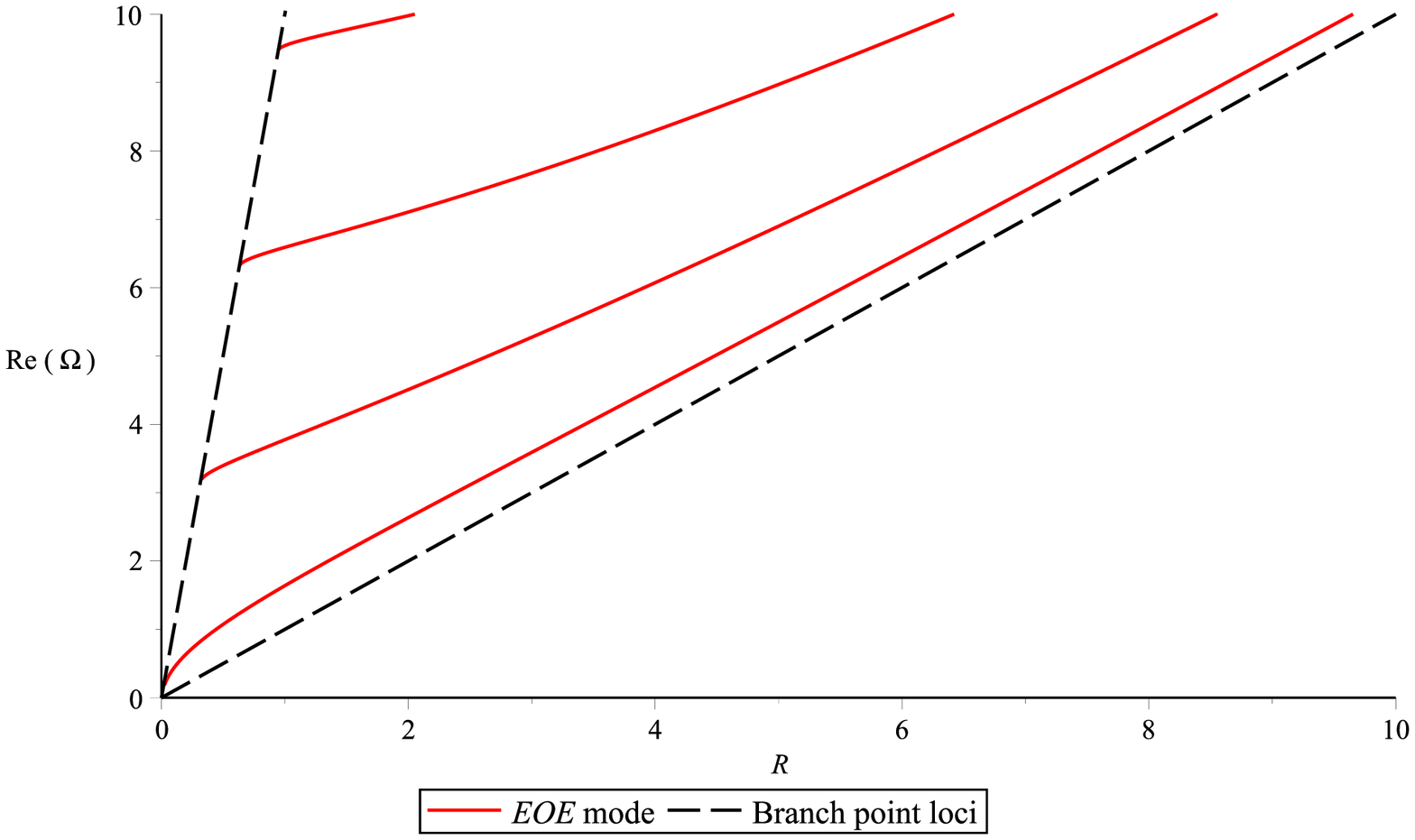}
    \caption{Location of category E poles and branch points with $\Omega'>0$ on real $\Omega$ axis for TE constant dielectric model with $\xi=\frac{1}{100}$ are located where a vertical line at fixed $R$ intersects any red loci and dashed loci respectively. The diagram exhibits no EEE or ELE modes with real frequency in this model. The indicated EOE mode loci follow from ${D}^{\TTE}_{\E\O\E,\kx,\ky}( \Omega',\xi)=0$ (\ref{DTE_EOE}). }
    \label{fig:TE_DIELECTRIC}
\end{figure}
\begin{figure}[h!]
    \centering
    \includegraphics[width=0.8\textwidth]{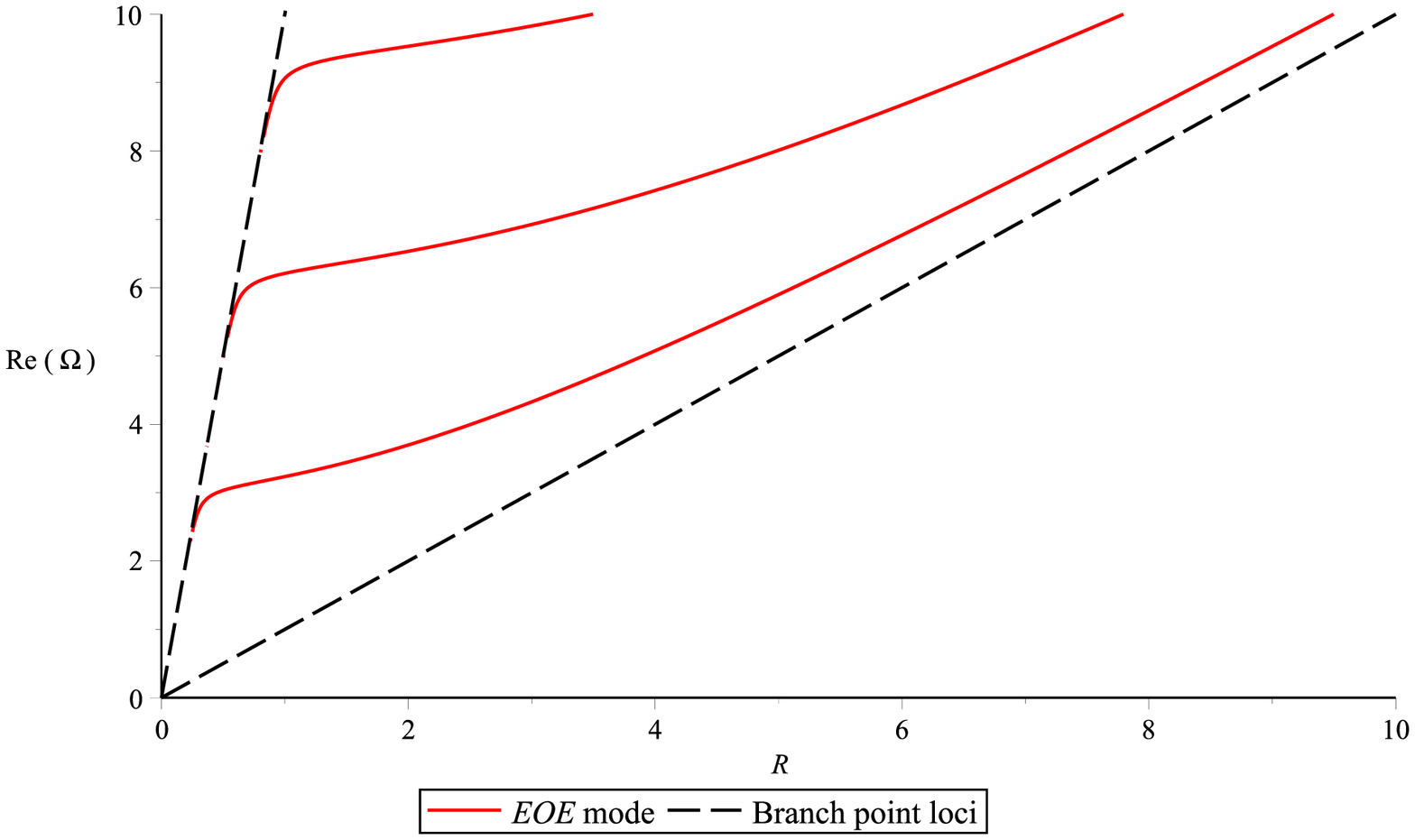}
    \caption{Location of category E poles  and branch points with $\Omega'>0$ on real $\Omega$ axis for TM constant dielectric model with $\xi=\frac{1}{100}$ are located where a vertical line at fixed $R$ intersects any red loci and dashed loci respectively. The diagram exhibits no EEE or ELE modes with real frequency in this model. The indicated EOE mode loci follow from ${D}^{\TTM}_{\E\O\E,\kx,\ky}( \Omega',\xi)=0$ (\ref{DTM_EOE}). }
    \label{fig:TM_DIELECTRIC}
\end{figure}
The nature of the category P and E real spectra as a function of $R$ for a fixed value of $\xi$ depends upon the behaviour of the relative
permittivity in regions I and III as a function of $\Omega'$ and $\xi$. Figures~\ref{fig:TE_PLASMA} and \ref{fig:TM_PLASMA} refer to the TE and TM modes respectively for a model with $\kr=1 - 30/(\Omega')^{2}$. Figures~\ref{fig:TE_DIELECTRIC} and \ref{fig:TM_DIELECTRIC} refer
similarly to a model with constant relative permittivity $\kr=\frac{1}{100}$. The parameter $\xi$ has been chosen to display generic features typical of each type of permittivity. It follows from (\ref{chizetadef}) that for the constant dielectric model, dielectrics with $\xi>1$ cannot sustain TE or TM EOE (dielectric waveguide) modes. Materials with $\xi<1$ are often referred to as epsilon-near-zero (ENZ) dielectrics \cite{alu2007epsilon}. In all figures, the dotted curves denote loci where $\chiI=0$ and $\chiII=0$. Red curves bounded by these dotted loci denote real loci where $D^{s}_{\E\O\E}$ is zero (waveguide modes). Blue curves below the dotted locus given by $\chiII=0$ denote real loci where $D^{s}_{\E\E\E}$ is zero (plasmon modes). The green locus in figure \ref{fig:TM_PLASMA} describes an allowable mode when $\chiII=0$ and $D^{\TTM}_{\E\L\E}=0$ are simultaneously satisfied. This occurs where the green curve intersects the dotted line $\chiII=0$ for a particular value of $R$. All these spectra are the limiting forms of the cavity spectra arising from the real zeroes of ${\cal D}^{s}_{\E\O\E},  {\cal D}^{s}_{\E\E\E}$ and $\left.{\cal D}^{s}_{\E\L\E}\right|_{\chiII=0}$ as the volume of the cavity tends to infinity. By contrast, the category P modes describing left and right propagating waves are not attenuated for large $|Z|$. However, one can construct from $T^{s,u}_{\O\O\O}(\Omega',\xi,R), \, T^{s,u}_{\O\E\O}(\Omega',\xi,R)$ and $T^{s,u}_{\O\L\O}(\Omega',\xi,R)$ their analytic continuations $\wt{T}^{s,u}_{\O\O\O}(\Omega,\xi,R), \,\wt{T}^{s,u}_{\O\E\O}(\Omega,\xi,R)$ and $\wt{T}^{s,u}_{\O\L\O}(\Omega,\xi,R)$ in the {\it right-half}  complex $\Omega$-plane by drawing cuts along the real axis from the branch points given by $\chiI(\Omega)=0$ and $\chiII(\Omega)=0$ to infinity. In particular, one finds that for fixed $\xi,R$ the complex function $\wt{T}^{s,u}_{\O\O\O}(\Omega,\xi,R)$ has simple poles located on the real axis at positions that coincide with the real zeroes of $D^{s}_{\E\O\E}, \,D^{s}_{\E\E\E}$ and $D^{s}_{\E\L\E}$. Although $\wt{T}^{s,u}_{\O\O\O}(\Omega,\xi,R)$ has complex poles for real $\Omega<0$, it only has simple poles and branch points (where $\chiI$ and $\chiII$ vanish) on the real axis for real $\Omega>0$ (by inspection). Furthermore, it has ``good'' analytic behaviour as $\rho\rightarrow\infty$ on the quadrant arcs $\man{A}^{\pm}_{\rho}$ where $\Omega=\rho e^{i\theta}$ for $0<\theta\leq\frac{\pi}{2}$ and $\frac{-\pi}{2}\leq\theta < 0$ respectively. More precisely, let
\begin{eqnarray*}
    \[\df \man{F}^{\TTE,u}(\Omega',\xi,R)\]^{-1} &=& \[ \frac{(\chiI+\chiII)^{2}e^{i(\chiI-\chiII)}}{4\chiI\chiII}\]T^{\TTE,u}_{\O\O\O}(\Omega',\xi,R)
\end{eqnarray*}
for $\text{Im}(\Omega)>0$, i.e.
\begin{eqnarray*}
    \man{F}^{\TTE,u}(\Omega',\xi, R) &=& 1 - \(\frac{\chiI - \chiII}{\chiI + \chiII}\)^{2}e^{2i\chiII},
\end{eqnarray*}
and
\begin{eqnarray*}
    \[\df \man{F}^{\TTM,u}(\Omega',\xi,R)\]^{-1} &=& \[ \frac{(\chiI+\kr\chiII)^{2}e^{i(\chiI-\chiII)}}{4\kr\chiI\chiII}\]T^{\TTM,u}_{\O\O\O}(\Omega',\xi,R)
\end{eqnarray*}
for $\text{Im}(\Omega)>0$, i.e.
\begin{eqnarray*}
    \man{F}^{\TTM,u}(\Omega',\xi, R) &=& 1 - \(\frac{\chiI - \kr\chiII}{\chiI + \kr\chiII}\)^{2}e^{2i\chiII},
\end{eqnarray*}
with continuation to $\wt{\man{F}}^{s,u}(\Omega,\xi, R)$ according to (\ref{AContinDef}).
Then $\partial_{\Omega}\ln\wt{\man{F}}^{s,u}(\Omega,\xi,R)$ has simple poles of unit residue at each isolated zero of $\wt{\man{F}}^{s,u}(\Omega,\xi,R)$ \, (pole of $\wt{T}^{s,u}_{\O\O\O}(\Omega,\xi,R)$) in the $\Omega$-plane. Since $\wt{\man{F}}^{s,u}$ has branch points and $\wt{T}^{s,u}_{\O\O\O}(\Omega,\xi,R)$ has been defined on the first Riemann sheet with cuts as
described above, in some circumstances real poles of $\wt{T}^{s,u}_{\O\O\O}(\Omega,\xi,R)$ may occur on top of the cuts, as illustrated in figures~\ref{fig:TE_PLASMA}-\ref{fig:TM_DIELECTRIC} for the waveguide modes or coincide with branch points, as in figure~\ref{fig:TM_PLASMA}. In these situations, the location of the poles at $\Omega=\Omega_{p}$ are displaced to $\Omega=\Omega_{p}-i\varepsilon$ and positive $\varepsilon\rightarrow 0$ after the residues have been extracted. As is clear from figures~\ref{fig:TE_PLASMA}-\ref{fig:TM_DIELECTRIC}, for fixed $\xi$, the location of all poles and branch points in the $\Omega$-plane move individually with different rates as a function of $R$. Furthermore, as $R$ increases, the number of poles between two branch points, in general, increases indefinitely.

\begin{figure}[h!]
\begin{center}
\setlength{\unitlength}{1cm}
\begin{picture}(13,13)
    \put(1,0){\psline[linewidth=1.5pt,arrowsize=4pt 4]{->}(0,0)(0,12)} 
    \put(0,6){\psline[linewidth=1.5pt,arrowsize=4pt 4]{->}(0,0)(11.8,0)} 
    \put(1.3,11){\psline[linewidth=1pt,arrowsize=3pt 3, linecolor=red]{->}(0,0)(0,-3)} 
    \put(1.3,11){\psline[linewidth=1pt,arrowsize=3pt 3, linecolor=red]{-}(0,0)(0,-10)}  
    \put(2.5,6){\circle*{0.2}} 
    \put(3.5,6){\circle*{0.2}} 
    \put(4.5,6){\psdot*[dotstyle=x, dotsize=6pt 5, linewidth=1.5pt](0,0)} 
    \put(8.5,6){\psdot*[dotstyle=x, dotsize=6pt 5, linewidth=1.5pt](0,0)} 
    \put(5.5,5.5){\circle*{0.2}} 
    \put(6.5,5.5){\circle*{0.2}} 
    \put(7.5,5.5){\circle*{0.2}} 
    \put(3,6){\psarcn[linewidth=1.02pt, linecolor=blue](0,0){1}{270}{90}} 
    \put(3,6){\psarcn[arrowsize=3pt 3, linewidth=1.02pt, linecolor=blue]{->}(0,0){1}{270}{140}} 
    \put(3,7){\psline[linewidth=1pt, linecolor=blue](0,0)(7,0)} 
    \put(3,7){\psline[linewidth=1pt, arrowsize=3pt 3, linecolor=blue]{->}(0,0)(3,0)} 
    \put(10,5){\psline[linewidth=1pt, linecolor=blue](0,0)(-7,0)} 
    \put(10,5){\psline[linewidth=1pt, arrowsize=3pt 3, linecolor=blue]{->}(0,0)(-4,0)} 
    \put(1.3,7){\psellipticarc[linewidth=1.1pt, linecolor=green]{-}(0,0)(8.7,4){0}{90}} 
    \put(1.3,7){\psellipticarc[linewidth=1.1pt, arrowsize=3pt 3, linecolor=green]{->}(0,0)(8.7,4){0}{30}} 
    \put(1.3,5){\psellipticarc[linewidth=1.1pt, linecolor=green]{-}(0,0)(8.7,4){270}{0}} 
    \put(1.3,5){\psellipticarc[linewidth=1.1pt, arrowsize=3pt 3, linecolor=green]{->}(0,0)(8.7,4){270}{330}} 
    \put(4.5,6.5){\pscoil[coilarm=0cm, coilaspect=0, coilheight=0.8, coilwidth=0.25,linewidth=1pt]{-}(0,0)(6.55,0)} 
    \put(4.5,6){\psline[linewidth=1pt](0,0)(0,0.5)} 
    \put(8.5,6){\pscoil[coilarm=0cm, coilaspect=0, coilheight=0.8, coilwidth=0.25,linewidth=1pt]{-}(0,0)(2.5,0)} 
    \put(6.85,10.35){\mbox{$\man{A}^{+}_{\rho}$}}
    \put(6.5,1.3){\mbox{$\man{A}^{-}_{\rho}$}}
    \put(4.3,5.4){\mbox{$\Omega'_{\tm{B1}}$}}
    \put(8.3,5.4){\mbox{\large $\Omega'_{\tm{B2}}$}}
    \put(5.5,5.55){\psline[arrowsize=2pt 2]{<->}(0,0)(0,0.45) } 
    \put(6.5,5.55){\psline[arrowsize=2pt 2]{<->}(0,0)(0,0.45) } 
    \put(7.5,5.55){\psline[arrowsize=2pt 2]{<->}(0,0)(0,0.45) } 
    \put(5.65,5.7){\mbox{$\small \varepsilon$}}
    \put(6.65,5.7){\mbox{$\small \varepsilon$}}
    \put(7.65,5.7){\mbox{$\small \varepsilon$}}
    \put(12,5.85){\mbox{\large $\Omega'$}}
    \put(0.85,12.1){\mbox{\large $\Omega''$}}
    \put(1.5,8.1){\mbox{$\Gamma''_{\rho}$}}
    \put(5.6,7.3){\mbox{$\Gamma'_{\rho}$}}
\end{picture}
\caption{Diagram showing the contour over which Cauchy's integral theorem is applied and the ``typical'' pole and branch point structure for the choice of $\kr(\Omega,R)$ models investigated. The green contour shows the quadrant arcs $\man{A}^{\pm}_{\rho}$ of radius $\rho$, the red contour $\Gamma''_{\rho}$ is along a finite segment of the $\Omega''$ axis and the blue contour $\Gamma'_{\rho}$ runs just above and below all cuts along a segment of the $\Omega'$ axis. The closed, counter-clockwise contour used in the Cauchy integral theorem in the text is $C=\Gamma''_{\rho}\cup\man{A}^{-}_{\rho}\cup \Gamma'_{\rho} \cup\man{A}^{+}_{\rho}$. The cuts associated with the branch points overlap on the real axis and poles may lie on these cuts. For typographic clarity and to facilitate computation of residues and discontinuities, these cuts and poles have been displaced slightly in the figure.}
\label{fig:ArgandD}
\end{center}
\end{figure}
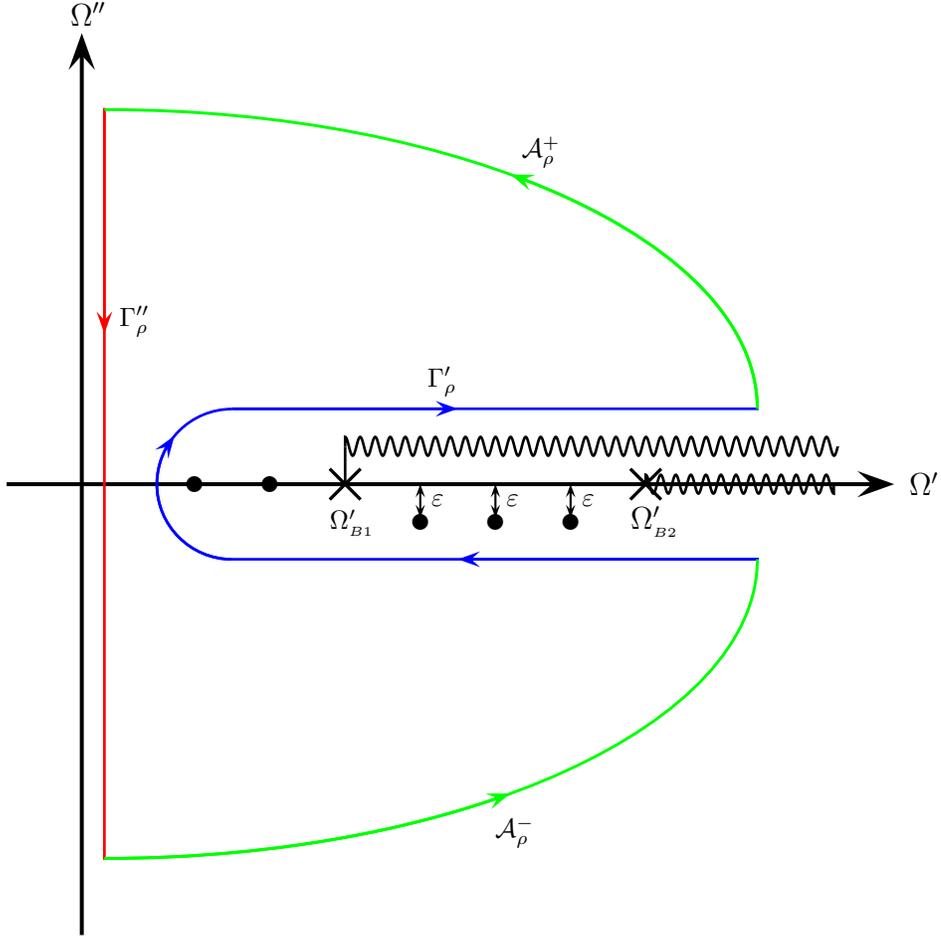
To construct the analogue of (\ref{pressureBoundDef}) based on a completeness relation analogous to (\ref{BoundedCR}) for discrete spectra in a cavity, one needs to sum over all modes in the open system with real spectra. Since the  modes with continuous spectra can only be normalized in a distributional sense, the mode superposition will rely on the analytic structure devised above to relate the mode summation with real frequencies to a finite expression with imaginary frequencies. For a fixed values of $\xi,R$ and some choice of $\kr(\Omega,R)$, a typical distribution of singularities of $\partial_{\Omega}\ln\wt{\man{F}}^{s,u}(\Omega,\xi,R)$ in the right-half $\Omega$-plane is displayed in figure~\ref{fig:ArgandD} where the poles between branch points at $\Omega'_{\tm{B1}}$ and $\Omega'_{\tm{B2}}$ have been displaced slightly into the lower half plane. The contour $C$ in figure~\ref{fig:ArgandD} encloses no singularities of any kind and hence one has from the Cauchy integral relation
\begin{eqnarray*}
    2\pi\int_{0}^{R_{0}} R\,dR \; \frac{1}{2\pi i}\int_{C} \Omega \pdiff{}{\Omega}\ln \wt{\man{F}}^{s,u}(\Omega,\xi,R)\, d\Omega &=& 0
\end{eqnarray*}
for finite $R_{0}$. Hence, by deforming $C=\Gamma''_{\rho}\cup\man{A}^{-}_{\rho}\cup\Gamma'_{\rho}\cup\man{A}^{+}_{\rho}$ to encircle the isolated poles, this yields:
\begin{eqnarray}\label{ContRel}
    \man{I}^{s}_{R_{0}}[-\Gamma''_{\rho}] &=& \man{I}^{s}_{R_{0}}[\man{A}^{-}_{\rho}] + \man{I}^{s}_{R_{0}}[\Gamma'_{\rho}] + \man{I}^{s}_{R_{0}}[\man{A}^{+}_{\rho}]
\end{eqnarray}
with $\man{I}^{s}_{R_{0}}[-\Gamma''_{\rho}]=-\man{I}^{s}_{R_{0}}[\Gamma''_{\rho}]$. The integrals over the various contours are given by
\begin{eqnarray*}
\begin{tabular}{rl}
    $\displaystyle\man{I}^{s}_{R_{0}}[-\Gamma''_{\rho}]$ & $\hspace{-0.3cm} \displaystyle \= \int_{0}^{R_{0}} \!\!\! R\,dR \!\int_{-\rho}^{\rho} \Omega''\pdiff{}{\Omega''}\ln \wt{\man{F}}^{s,u}(i\Omega'',\xi,R)\, d\Omega''$ \\
    $\displaystyle\man{I}^{s}_{R_{0}}[\Gamma'_{\rho}]$ & $\hspace{-0.3cm} \displaystyle \= 2\pi\!\!\int_{0}^{R_{0}} \!\!\! R\,dR \! \( -\!\!\!\!\!\sum_{p=1}^{N^{s}(\xi,R)}\!\!\! \Omega_{p}^{s}(\xi,R) + \frac{1}{\pi}\!\int_{\Omega'_{\tm{B1}}}^{\rho} \!\!\!\!\Omega'\pdiff{}{\Omega'}\,\text{Im}\!\[\df \! \ln \wt{\man{F}}^{s,u}(\Omega',\xi,R)\] \! d\Omega' \! \)$
\end{tabular}
\end{eqnarray*}
with $N^{s}(\xi,R)$ denoting the total number of poles that contribute for each $R$ in the range $0$ to $R_{0}$ and
\begin{eqnarray*}
    \man{I}^{s}_{R_{0}}[\man{A}^{\pm}_{\rho}] &=& 2\pi\int_{0}^{R_{0}} \! R\,dR \; \frac{1}{2\pi i}\int_{\man{A}^{\pm}_{\rho}} \Omega\pdiff{}{\Omega}\ln \wt{\man{F}}^{s,u}(\Omega,\xi,R)\, d\Omega
\end{eqnarray*}
are the contributions to the contour integral from the quadrant arcs of radius $\rho$ in the regions $\text{Im}(\Omega)>0$ and $\text{Im}(\Omega)<0$ respectively. Furthermore, the contribution to $\man{I}^{s}_{R_{0}}[-\Gamma''_{\rho}]$ from the integrands with $\Omega''\geq 0$ and $\Omega''\leq0$ can each be written in terms of $\man{F}^{s,u}(i\Omega'',\xi,R)$ according to (\ref{AContinDef}), to obtain
\begin{eqnarray}\label{dispInt}
    \man{I}^{s}_{R_{0}}[-\Gamma''_{\rho}] &=& 2\int_{0}^{R_{0}} \! R\,dR \int_{0}^{\rho} \Omega''\pdiff{}{\Omega''}\Re\[ \df \ln \man{F}^{s,u}(i\Omega'',\xi,R)\]\, d\Omega'' .
\end{eqnarray}
In particular, the function $\man{F}^{s,u}(i\Omega'',\xi,R)$ is real-valued provided $\kr(i\Omega'',\xi)$ is also real-valued. This holds for the simple dielectric model and plasma model we are considering, yielding
\begin{eqnarray}\label{nondispInt}
    \man{I}^{s}_{R_{0}}[-\Gamma''_{\rho}] &=& 2\int_{0}^{R_{0}} \! R\,dR \int_{0}^{\rho} \Omega''\pdiff{}{\Omega''}\ln \man{F}^{s,u}(i\Omega'',\xi,R)\, d\Omega'' .
\end{eqnarray}
Clearly, applications to other dielectric models will require detailed analysis of their complex analytic properties in $\Omega$.
Furthermore, the summations in (\ref{ContRel}) are finite. If one defines the \textit{real valued function} $\delta^{s,u}(\Omega',\xi,R)$ by
\begin{eqnarray*}
    \delta^{s,u}(\Omega',\xi,R) &=& \frac{1}{2i}\ln\( \frac{T^{s,u}_{\O\O\O}(\Omega',\xi,R)}{\conj{T^{s,u}_{\O\O\O}}(\Omega',\xi,R)} \) \=
    -\frac{1}{2i}\[ \ln \man{F}^{s,u}(\Omega',\xi,R) - \ln \conj{\man{F}^{s,u}}(\Omega',\xi,R)\df \] \\
    &=& -\text{Im}\( \df \ln \man{F}^{s,u}(\Omega',\xi,R) \),
\end{eqnarray*}
one may write the integral over $\Gamma'_{\rho}$ as
\begin{eqnarray}\label{IGammaPRho}
    \man{I}^{s}_{R_{0}}[\Gamma'_{\rho}] &=& -2\pi\int_{0}^{R_{0}} \!\! R\,dR \( \sum_{p=1}^{N^{s}(\xi,R)} \Omega_{p}^{s}(\xi,R) + \frac{1}{\pi}\int_{\Omega'_{\tm{B1}}}^{\rho} \Omega'\pdiff{}{\Omega'}\delta^{s,u}(\Omega',\xi,R)\, d\Omega'\)
\end{eqnarray}
since $\wt{\man{F}}=\man{F}$ just above all cuts along the real axis. This may be interpreted as a continuous real spectral summation associated with a superposition of all $s$-modes with $R\leq R_{0}$, $\Omega'<\rho$ characterized by $\xi$ for open media in regions I and III and a vacuum in region II.

The behaviour of $\man{I}^{s}_{R_{0}}[\Gamma'_{\rho}]$ as $R_{0}\rightarrow\infty$ depends upon the behaviour of $N^{s}(\xi,R)$,  $\delta^{s,u}(\Omega',\xi,R)$ and $\Omega_{p}^{s}(\xi,R)$ as $R\rightarrow\infty$. From (\ref{ContRel}), this in turn depends on how the integrand of the double integral of $\man{I}^{s}_{R_{0}}[-\Gamma''_{\rho}]$ behaves in this limit. Remarkably, for a large class of permittivities $\kr(\Omega,\xi)$, it is finite in this limit, implying that any divergences in the terms in (\ref{IGammaPRho}) cancel. Furthermore, for permittivities in this class, the contributions $\man{I}^{s}_{R_{0}}[\man{A}^{\pm}_{\rho}]\rightarrow 0$ as $\rho\rightarrow\infty$. It remains to endow the integrals on the left (or right) of (\ref{ContRel}) in these limits with physical content.

Consider a system with a medium of a constant permittivity $\ep{0}\kr_{0}$ in regions I and III. As $\kr_{0}\rightarrow \infty$, the electromagnetic fields in these regions will attenuate to zero and one emulates a system with an open finite width planar vacuum slab that partitions a perfectly conducting open medium. No non-zero propagating modes can survive this limit and only standing waves inside the vacuum region satisfying perfectly conducting junction conditions at the interfaces can contribute to a quantum stress on either interface. The magnitude of this stress should coincide with that computed by enclosing a finite volume vacuum domain in a perfectly conducting cuboid cavity and letting two opposite faces of the cuboid expand to infinity, whilst keeping the third dimension $\Lz$ of the cuboid finite. The standard computation of this particular Casimir stress requires some type of regularization process to extract a quantum induced interaction energy per unit area $\man{U}^{s}_{\tt{Casimir}}$ of magnitude $\pi^{2}\hbar c/1440\Lz^{3}$ contributed by all modes of type $s$ in this situation. By contrast, with $\xi\equiv\kr_{0} \in \real$, the limit
\begin{eqnarray*}
    \lim_{\kr_{0}\rightarrow \infty} \Omega''\pdiff{}{\Omega''}\ln \man{F}^{s,u}(i\Omega'',\kr_{0},R) &=&
    \frac{2(\Omega'')^{2}}{\sqrt{R^{2}+(\Omega'')^{2}}\(e^{2\sqrt{R^{2}+(\Omega'')^{2}}} - 1 \) }
\end{eqnarray*}
yields the dimensionless integral
\begin{eqnarray*}
    \man{I}^{s}_{\infty} &\equiv& \lim_{\kr_{0}\rightarrow\infty}\man{I}_{\infty}^{s}[-\Gamma''_{\infty}] \= 2\int_{0}^{\infty} R\,dR \int_{0}^{\infty} \Omega'' \pdiff{}{\Omega''} \ln \man{F}^{s,u}(i\Omega'',\infty,R)
    \, d\Omega'' \= \frac{\pi^{4}}{180}
\end{eqnarray*}
using (\ref{nondispInt}). For an open system with a slot of width $\Lz$, the only geometric scales available are $\Lz$ and the time it takes light to traverse this width in the vacuum. The quantum induced stress normal to the $z$-axis is then
\begin{eqnarray*}
    \man{P}^{s}_{\tt{Casimir}}(\Lz) &=& -\pdiff{}{\Lz} \man{U}^{s}_{\tt{Casimir}}(\Lz)
\end{eqnarray*}
where the interaction energy per unit area contributed by all modes of type $s$ is:
\begin{eqnarray*}
    \man{U}^{s}_{\tt{Casimir}}(\Lz) &=& -\frac{\hbar c}{\Lz^{3}}W_{0}^{s}\man{I}^{s}_{\infty}
\end{eqnarray*}
for some dimensionless constant $W_{0}^{s}$. Thus, $\man{P}^{s}_{\tt{Casimir}}(\Lz)$ yields the experimentally confirmed Casimir pressure if the normalization constant $W_{0}^{s}=1/8\pi^{2}$. If $\man{I}^{s}_{R_{0}}[\Gamma'_{\rho}]$ and $\man{I}^{s}_{R_{0}}[\Gamma''_{\rho}]$ remain finite for $R_{0}\rightarrow\infty$ and $\rho\rightarrow\infty$ for the system under consideration and the integrals over the quadrant arcs $\man{I}^{s}_{R_{0}}[\man{A}^{\pm}_{\rho}]\rightarrow 0$ in this limit, then from (\ref{ContRel}), one has
\begin{eqnarray*}
    \man{I}^{s}_{\infty}[-\Gamma''_{\infty}] &=& \man{I}^{s}_{\infty}[\Gamma'_{\infty}].
\end{eqnarray*}
If one multiplies both sides of this with $-\hbar c/8\pi^{2}\Lz^{3}$, each side may be identified with a quantum induced interaction energy per unit area:
\begin{eqnarray}
 \label{LifForm} \man{U}^{s}(\Lz,\xi) &=& -\frac{\hbar c}{4\pi^{2}\Lz^{3}}\int_{0}^{\infty} \!\! R\,dR \int_{0}^{\infty}\Omega''\pdiff{}{\Omega''}\ln \man{F}^{s,u}(i\Omega'',\xi,R)\, d\Omega'' \\
  \nonumber  &=& -\frac{\hbar c}{4\pi \Lz^{3}}\int_{0}^{\infty}\!\! R\,dR \[ \sum_{p=1}^{N^{s}(\xi,R)} \Omega_{p}^{s}(\xi,R) + \frac{1}{\pi}\int_{\Omega'_{\tm{B1}}}^{\infty}\!\!\Omega'\pdiff{}{\Omega'} \delta^{s,u}(\Omega',\xi,R) \, d\Omega'\]
\end{eqnarray}
The interaction energy per unit area for all modes of type $s$ given by (\ref{LifForm}) coincides with that found by Lifshitz \cite{Bordag,Lifshitz2,nesterenko2012lifshitz,bordag2012electromagnetic}. Hence, the quantum induced normal stress on either interface is a pressure of magnitude $\sum_{s}\man{P}^{s}$ where
\begin{eqnarray*}
    \footnotesize
\begin{tabular}{ll}
    $\man{P}^{s}(\Lz,\xi)$ & $ \displaystyle\hspace{-0.35cm} \= \left| \frac{\hbar c}{4\pi^{2}}\pdiff{}{\Lz}\( \frac{1}{\Lz^{3}}\int_{0}^{\infty} \!\! R\,dR \int_{0}^{\infty}\Omega''\pdiff{}{\Omega''}\ln \man{F}^{s,u}(i\Omega'',\xi,R)\, d\Omega''\) \right|$ \\
    & \\
    & $\displaystyle \hspace{-0.35cm}\= \left| \frac{\hbar c}{4\pi}\pdiff{}{\Lz}\( \frac{1}{\Lz^{3}}\int_{0}^{\infty}\!\! R\,dR \[ \sum_{p=1}^{N^{s}(\xi,R)} \Omega_{p}^{s}(\xi,R) + \frac{1}{\pi}\int_{\Omega'_{\tm{B1}}}^{\infty}\!\!\Omega'\pdiff{}{\Omega'} \delta^{s,u}(\Omega',\xi,R) \, d\Omega'\] \) \right|$
\end{tabular}
\end{eqnarray*}
Furthermore, in many models of relevance, an integration by parts yields
\begin{eqnarray*}
    \int_{0}^{\infty} \Omega''\pdiff{}{\Omega''}\ln \man{F}^{s,u}(i\Omega'',\xi,R)\, d\Omega'' &=& -\int_{0}^{\infty} \ln
    \man{F}^{s,u}(i\Omega'',\xi,R)\, d\Omega''.
\end{eqnarray*}
If one chooses to write the general interaction energy per unit area for all modes of type $s$ as
\begin{eqnarray*}
    \man{U}^{s}(\Lz,\xi) &=& -\frac{\hbar c}{4\pi \Lz^{3}}\int_{-\infty}^{\infty} \Omega' d\man{N}^{s}(\Omega',\Lz,\xi) ,
\end{eqnarray*}
then the associated density of states measure $d\man{N}^{s}$ takes on a distributional form with
\begin{eqnarray*}
\footnotesize
\begin{tabular}{ll}
    $\displaystyle \pdiff{\man{N}^{s}}{\Omega'}(\Omega',\Lz,\xi)$ & $\displaystyle \hspace{-0.35cm}\= \int_{0}^{\infty} R\, dR \[ \sum_{p=1}^{N^{s}(\xi,R)}
    \frac{\delta_{D}\(\Omega'-\Omega_{p}^{s}(\xi,R)\)}{\Omega'} + \frac{1}{\pi}\man{H}(\Omega'-\Omega'_{\tm{B1}})\pdiff{}{\Omega'} \delta^{s,u}(\Omega',\xi,R)
    \,\]$
\end{tabular}
\end{eqnarray*}
in terms of the one dimensional Dirac distribution $\delta_{D}$ and the Heaviside function $\man{H}$.

\section{Conclusions}
This article has demonstrated how a number of different viewpoints on the genesis of the Lifshitz formula for a pair of separated planar homogeneous dielectric half-spaces may be unified by exploiting the analytic structure of certain functions derived from general solutions of the macroscopic Maxwell equations. By devising a state labelling scheme for stationary electromagnetic fields in piecewise continuous media it is possible to describe the physical content of mode summations that arise in the theory of electromagnetic quantum fluctuations for both open and closed systems. For closed systems attention has been drawn to discrete affine-linear modes in addition to those referred to as discrete waveguide and plasmon modes. For our open system, the full spectrum is continuous and mode identification can be usefully described in terms of complex analytic continuations between complex category E and category P amplitudes and associated local and global dispersion relations in media.  In this framework the locations of poles and branch points in the complex frequency plane (associated with continuous modes)  vary as functions of continuous (transverse) wave momenta.  The category E affine-linear states can arise as particular poles that coincide in location with certain branch points in particular situations. We have illustrated how the Cauchy integral formula can be used in the right-half complex angular frequency plane to extract a real mode summation (thereby determining a density of states) that may be identified up to scaling with a finite interaction energy-density equal to that given by the Lifshitz formula. By fixing the scaling by demanding agreement with the experimentally verified Casimir attraction between conducting plates one has  expressions for quantum stresses in the dielectric half-spaces in terms of permittivities with either real or pure imaginary frequencies.

The mode labelling scheme relies heavily on the separability of the Helmholtz equation (and its boundary and interface conditions) in the planar Lifshitz geometry.  Many other geometries are amenable to this scheme provided the system has interfaces and boundaries that coincide with parts of constant coordinate surfaces in $\real^{3}$ and  thereby admit separability. The methodology may also be generalized to non-simply connected domains by including additional modes (TEM) in the Hodge-de-Rham decomposition. However, any attempt to generalize along these lines for open systems demands the explicit determination of a basis of solutions to the macroscopic Maxwell equations compatible with the prescribed interface conditions with possibly dispersive and inhomogeneous media. If this can be achieved the role of analyticity in complex frequency in extending Lifshitz theory to address some of the unsolved problems outlined in the introduction may prove fruitful.

\section*{Acknowledgements}
As members of the ALPHA-X collaboration and the Cockcroft Institute
of Accelerator Science and Technology the authors are grateful for support from EPSRC (EP/J018171/) and STFC (ST/G008248/1).

\appendix
\section{Modes in Closed Domains}\label{BoundAppendix}
Following (\ref{chizetadef}) and the discussion in section~\ref{CavitySect}, the cavity has geometric aspect ratio defined by the parameters $\underline{\lambda}$ and material configuration defined by $\kr^{\I}(\Omega,\xi)=\kr^{\III}(\Omega,\xi)\equiv \kr(\Omega,\xi)$ and $\kr^{\II}=1$ with $\kr$ dependent upon a single material constant $\xi$.

\subsection{$\man{N}$=OOO Cavity Discrete Spectrum Generators}
From the interface and perfectly conducting boundary conditions discussed in section~\ref{CavitySect}, one finds, for the TE and TM modes respectively, the homogeneous algebraic systems:
\begin{eqnarray*}
 \scriptsize
\begin{tabular}{l}
    $\( \!\!\! \begin{array}{cccccc}
            1 & 0 & -1 & 0 & 0 & 0 \\
            0 & 0 & \cos(\chiII) & \sin(\chiII) & -\cos(\chiI) & -\sin(\chiI) \\
            0 & \chiI & 0 & -\chiII & 0 & 0 \\
            0 & 0 & \chiII\sin(\chiII) & -\chiII\cos(\chiII) & -\chiI\sin(\chiI) & \chiI\cos(\chiI) \\
            \cos(\lambda\chiI) & -\sin(\lambda\chiI) & 0 & 0 & 0 & 0 \\
            0 & 0 & 0 & 0 & \cos(\eta_{\O}) & \sin(\eta_{\O}) \\
        \end{array} \!\!\! \) \!\! \( \!\!\!  \begin{array}{c}
                                \MMD{P}{\TTE}{\I}{\Omega}{\O} \\
                                \MMD{Q}{\TTE}{\I}{\Omega}{\O} \\
                                \MMD{P}{\TTE}{\II}{\Omega}{\O} \\
                                \MMD{Q}{\TTE}{\II}{\Omega}{\O} \\
                                \MMD{P}{\TTE}{\III}{\Omega}{\O} \\
                                \MMD{Q}{\TTE}{\III}{\Omega}{\O} \\
                            \end{array} \!\!\! \) \!\! \= 0$   \\
                            \\
    $\( \!\!\! \begin{array}{cccccc}
            0 & \chiI & 0 & -\chiII & 0 & 0 \\
            0 & 0 & \chiII\sin(\chiII) & -\chiII\cos(\chiII) & -\chiI\sin(\chiI) & \chiI\cos(\chiI) \\
            \kr & 0 & -1 & 0 & 0 & 0 \\
            0 & 0 & \cos(\chiII) & \sin(\chiII) & -\kr\cos(\chiI) & -\kr\sin(\chiI) \\
            \sin(\lambda\chiI) & \cos(\lambda\chiI) & 0 & 0 & 0 & 0 \\
            0 & 0 & 0 & 0 & \sin(\eta_{\O}) & -\cos(\eta_{\O}) \\
        \end{array} \!\!\! \) \!\! \(  \!\!\!  \begin{array}{c}
                                \MMD{P}{\TTM}{\I}{\Omega}{\O} \\
                                \MMD{Q}{\TTM}{\I}{\Omega}{\O} \\
                                \MMD{P}{\TTM}{\II}{\Omega}{\O} \\
                                \MMD{Q}{\TTM}{\II}{\Omega}{\O} \\
                                \MMD{P}{\TTM}{\III}{\Omega}{\O} \\
                                \MMD{Q}{\TTM}{\III}{\Omega}{\O} \\
                            \end{array} \! \!\! \) \!\!\= 0  $
\end{tabular}
\end{eqnarray*}\normalsize
where $\eta_{\O}=(1+\lambda)\chiI$. These have non-trivial solutions provided the following determinants vanish:
\begin{eqnarray*}
 \footnotesize
\begin{tabular}{ll}
    ${\cal D}^{\TTE}_{\O\O\O,n_x,n_y}( \Omega,\underline{\lambda},\xi)$ & $\hspace{-0.35cm}\= \sin(\chiII)\( \df \chiII^{2}\sin^{2}(\lambda\chiI) -
    \chiI^{2}\cos^{2}(\lambda\chiI) \) - \chiI\chiII\cos(\chiII)\sin(2\lambda\chiI)$ \\
    ${\cal D}^{\TTM}_{\O\O\O,n_x,n_y}( \Omega,\underline{\lambda},\xi)$ & $\hspace{-0.35cm}\= \sin(\chiII)\( \df \kr^{2}\chiII^{2}\cos^{2}(\lambda\chiI) -
    \chiI^{2}\sin^{2}(\lambda\chiI) \) - \kr\chiI\chiII\cos(\chiII)\sin(2\lambda\chiI)$,
\end{tabular}
\end{eqnarray*}
thereby determining (up to scaling) a class of allowed modes with real discrete spectra.

\subsection{$\man{N}$=OEO Cavity Discrete Spectrum Generators}
Similarly, for this configuration, one has, for the TE and TM modes respectively, the algebraic systems:
\begin{eqnarray*}
 \scriptsize
\begin{tabular}{l}
    $\( \!\!\! \begin{array}{cccccc}
            1 & 0 & -1 & 0 & 0 & 0 \\
            0 & 0 & \cosh(\zetaII) & \sinh(\zetaII) & -\cos(\chiI) & -\sin(\chiI) \\
            0 & \chiI & 0 & -\zetaII & 0 & 0 \\
            0 & 0 & \zetaII\sinh(\zetaII) & \zetaII\cosh(\zetaII) & \chiI\sin(\chiI) & -\chiI\cos(\chiI) \\
            \cos(\lambda\chiI) & -\sin(\lambda\chiI) & 0 & 0 & 0 & 0 \\
            0 & 0 & 0 & 0 & \cos(\eta_{\O}) & \sin(\eta_{\O}) \\
        \end{array} \!\!\!\) \!\! \( \!\!\!  \begin{array}{c}
                                \MMD{P}{\TTE}{\I}{\Omega}{\O} \\
                                \MMD{Q}{\TTE}{\I}{\Omega}{\O} \\
                                \MMD{P}{\TTE}{\II}{\Omega}{\E} \\
                                \MMD{Q}{\TTE}{\II}{\Omega}{\E} \\
                                \MMD{P}{\TTE}{\III}{\Omega}{\O} \\
                                \MMD{Q}{\TTE}{\III}{\Omega}{\O} \\
                            \end{array} \!\!\!\) \!\!\= 0$  \\
                             \\
    $\( \!\!\! \begin{array}{cccccc}
            0 & \chiI & 0 & -\zetaII & 0 & 0 \\
            0 & 0 & \zetaII\sinh(\zetaII) & \zetaII\cosh(\zetaII) & \chiI\sin(\chiI) & -\chiI\cos(\chiI) \\
            \kr & 0 & -1 & 0 & 0 & 0 \\
            0 & 0 & \cosh(\zetaII) & \sinh(\zetaII) & -\kr\cos(\chiI) & -\kr\sin(\chiI) \\
            \sin(\lambda\chiI) & \cos(\lambda\chiI) & 0 & 0 & 0 & 0 \\
            0 & 0 & 0 & 0 & \sin(\eta_{\O}) & -\cos(\eta_{\O}) \\
        \end{array} \!\!\! \) \!\! \( \!\!\!  \begin{array}{c}
                                \MMD{P}{\TTM}{\I}{\Omega}{\O} \\
                                \MMD{Q}{\TTM}{\I}{\Omega}{\O} \\
                                \MMD{P}{\TTM}{\II}{\Omega}{\E} \\
                                \MMD{Q}{\TTM}{\II}{\Omega}{\E} \\
                                \MMD{P}{\TTM}{\III}{\Omega}{\O} \\
                                \MMD{Q}{\TTM}{\III}{\Omega}{\O} \\
                            \end{array} \!\!\! \) \!\!\= 0 $
\end{tabular}
\end{eqnarray*}
where $\eta_{\O}=(1+\lambda)\chiI$ with corresponding determinants:
\begin{eqnarray*}
 \footnotesize
\begin{tabular}{ll}
    ${\cal D}^{\TTE}_{\O\E\O,n_x,n_y}( \Omega,\underline{\lambda},\xi)$ & $\hspace{-0.35cm}\= \sinh(\zetaII)\( \df \zetaII^{2}\sin^{2}(\lambda\chiI) +
    \chiI^{2}\cos^{2}(\lambda\chiI) \) + \chiI\zetaII\cosh(\zetaII)\sin(2\lambda\chiI)$ \\
    ${\cal D}^{\TTM}_{\O\E\O,n_x,n_y}( \Omega,\underline{\lambda},\xi)$ & $\hspace{-0.35cm}\= \sinh(\zetaII)\( \df  \kr^{2}\zetaII^{2}\cos^{2}(\lambda\chiI) +
    \chiI^{2}\sin^{2}(\lambda\chiI) \) - \kr\zetaI\zetaII\cosh(\zetaII)\sin(2\lambda\chiI)$.
\end{tabular}
\end{eqnarray*}

\subsection{$\man{N}$=OLO Cavity Discrete  Spectrum Generators}
For this configuration, $\chiII=0$ (i.e. $\Omega=R$) and the systems for the TE and TM modes respectively are:
\begin{eqnarray*}
 \small
\begin{tabular}{l}
   $\( \!\!\! \begin{array}{cccccc}
            1 & 0 & -1 & 0 & 0 & 0 \\
            0 & 0 & 1 & 1 & -\cos(\chiI) & -\sin(\chiI) \\
            0 & \chiI & 0 & -1 & 0 & 0 \\
            0 & 0 & 0 & 1 & \chiI\sin(\chiI) & -\chiI\cos(\chiI) \\
            \cos(\lambda\chiI) & -\sin(\lambda\chiI) & 0 & 0 & 0 & 0 \\
            0 & 0 & 0 & 0 & \cos(\[1+\lambda\]\chiI) & \sin(\[1+\lambda\]\chiI) \\
        \end{array} \!\!\!\) \!\! \(\!\!\!   \begin{array}{c}
                                \MMD{P}{\TTE}{\I}{\Omega}{\O} \\
                                \MMD{Q}{\TTE}{\I}{\Omega}{\O} \\
                                \MMD{P}{\TTE}{\II}{\Omega}{\L} \\
                                \MMD{Q}{\TTE}{\II}{\Omega}{\L} \\
                                \MMD{P}{\TTE}{\III}{\Omega}{\O} \\
                                \MMD{Q}{\TTE}{\III}{\Omega}{\O} \\
                            \end{array} \!\!\!\) \= 0$ \\
                              \\
   $\( \!\!\! \begin{array}{cccccc}
            0 & \chiI & 0 & -1 & 0 & 0 \\
            0 & 0 & 0 & 1 & \chiI\sin(\chiI) & -\chiI\cos(\chiI) \\
            \kr & 0 & -1 & 0 & 0 & 0 \\
            0 & 0 & 1 & 1 & -\kr\cos(\chiI) & -\kr\sin(\chiI) \\
            \sin(\lambda\chiI) & \cos(\lambda\chiI) & 0 & 0 & 0 & 0 \\
            0 & 0 & 0 & 0 & \sin(\[1+\lambda\]\chiI) & -\cos(\[1+\lambda\]\chiI) \\
        \end{array} \!\!\! \) \!\! \(  \!\!\! \begin{array}{c}
                                \MMD{P}{\TTM}{\I}{\Omega}{\O} \\
                                \MMD{Q}{\TTM}{\I}{\Omega}{\O} \\
                                \MMD{P}{\TTM}{\II}{\Omega}{\L} \\
                                \MMD{Q}{\TTM}{\II}{\Omega}{\L} \\
                                \MMD{P}{\TTM}{\III}{\Omega}{\O} \\
                                \MMD{Q}{\TTM}{\III}{\Omega}{\O} \\
                            \end{array} \!\!\! \) \= 0$
\end{tabular}
\end{eqnarray*}
with corresponding determinants:
\begin{eqnarray*}
 \small
\begin{tabular}{ll}
    ${\cal D}^{\TTE}_{\O\L\O,n_x,n_y}( \Omega,\underline{\lambda},\xi)$ & $\hspace{-0.35cm}\= \chiI\cos(\lambda\chiI)\( \df2\sin(\lambda\chiI) + \chiI\cos(\lambda\chiI) \)$ \\
    ${\cal D}^{\TTM}_{\O\L\O,n_x,n_y}( \Omega,\underline{\lambda},\xi)$ & $\hspace{-0.35cm}\= \chiI\sin(\lambda\chiI)\( \df 2\kr\cos(\lambda\chiI) - \chiI\sin(\lambda\chiI) \)$ .
\end{tabular}
\end{eqnarray*}
In the constant dielectric model with $\kr=\xi$, the vanishing of these determinants is reminiscent of a class of Diophantine equations. Since $\nx,\ny$ are integers and the trigonometric functions have periods proportional to the integers, they will not in general have solutions for arbitrary real $\xi$. However, there may exist many roots for particular values of such $\xi$. For example, for the TE configuration, the solutions arising from $\cos(\lambda\chiI)=\cos(\lambda R\sqrt{\xi-1})=0$ imply $\lambda R\sqrt{\xi-1} \= \(\ell + \frac{1}{2}\)\pi$ for some integer $\ell\geq 0$. From (\ref{rRdef}), it follows that with $\lambda_{1}=\lambda_{2}=\lambda=1$:
\begin{eqnarray*}
    \nx^{2} + \ny^{2} \= \frac{\(\ell + \frac{1}{2}\)^{2}}{\xi-1},
\end{eqnarray*}
in terms of the three integers $(\nx,\ny,\ell)$. For a particular choice of $\xi$, there may or may not be solutions. For example, with
$\xi=\frac{9}{8}$, one has
\begin{eqnarray*}
    \nx^{2} + \ny^{2} \= 8\(\ell + \frac{1}{2}\)^{2}
\end{eqnarray*}
which has solutions $\nx=1,\ny=1$ for $\ell=0$ and $\nx=3,\ny=3$ for $\ell=1$ and so on. But with $\xi=2$, there can be no solutions since $\(\ell + \frac{1}{2}\)^{2}$ is non-integer.

\subsection{$\man{N}$=EOE Cavity Discrete  Spectrum Generators}
For this configuration, for the TE and TM modes respectively, the systems are:
\begin{eqnarray*}
 \scriptsize
\begin{tabular}{l}
    $\( \!\!\! \begin{array}{cccccc}
            1 & 0 & -1 & 0 & 0 & 0 \\
            0 & 0 & \cos(\chiII) & \sin(\chiII) & -\cosh(\zetaI) & -\sinh(\zetaI) \\
            0 & \zetaI & 0 & -\chiII & 0 & 0 \\
            0 & 0 & \chiII\sin(\chiII) & -\chiII\cos(\chiII) & \zetaI\sinh(\zetaI) & \zetaI\cosh(\zetaI) \\
            \cosh(\lambda\zetaI) & -\sinh(\lambda\zetaI) & 0 & 0 & 0 & 0 \\
            0 & 0 & 0 & 0 & \cosh(\eta_{\E}) & \sinh(\eta_{\E}) \\
        \end{array} \!\!\! \) \!\! \(\!\!\!   \begin{array}{c}
                                \MMD{P}{\TTE}{\I}{\Omega}{\E} \\
                                \MMD{Q}{\TTE}{\I}{\Omega}{\E} \\
                                \MMD{P}{\TTE}{\II}{\Omega}{\O} \\
                                \MMD{Q}{\TTE}{\II}{\Omega}{\O} \\
                                \MMD{P}{\TTE}{\III}{\Omega}{\E} \\
                                \MMD{Q}{\TTE}{\III}{\Omega}{\E} \\
                            \end{array} \!\!\!\) \!\!\= 0$ \\
                            \\
    $\( \!\!\! \begin{array}{cccccc}
            0 & \zetaI & 0 & -\chiII & 0 & 0 \\
            0 & 0 & \chiII\sin(\chiII) & -\chiII\cos(\chiII) & \zetaI\sinh(\zetaI) & \zetaI\cosh(\zetaI) \\
            \kr & 0 & -1 & 0 & 0 & 0 \\
            0 & 0 & \cos(\chiII) & \sin(\chiII) & -\kr\cosh(\zetaI) & -\kr\sinh(\zetaI) \\
            \sinh(\lambda\zetaI) & -\cosh(\lambda\zetaI) & 0 & 0 & 0 & 0 \\
            0 & 0 & 0 & 0 & \sinh(\eta_{\E}) & \cosh(\eta_{\E}) \\
        \end{array} \!\!\!\) \!\! \(\!\!\!   \begin{array}{c}
                                \MMD{P}{\TTM}{\I}{\Omega}{\E} \\
                                \MMD{Q}{\TTM}{\I}{\Omega}{\E} \\
                                \MMD{P}{\TTM}{\II}{\Omega}{\O} \\
                                \MMD{Q}{\TTM}{\II}{\Omega}{\O} \\
                                \MMD{P}{\TTM}{\III}{\Omega}{\E} \\
                                \MMD{Q}{\TTM}{\III}{\Omega}{\E} \\
                            \end{array} \!\!\!\) \!\!\= 0$
\end{tabular}
\end{eqnarray*}
where $\eta_{\E}=(1+\lambda)\zetaI$, with corresponding determinants:
\begin{eqnarray*}
 \footnotesize
\begin{tabular}{ll}
    ${\cal D}^{\TTE}_{\E\O\E,n_x,n_y}( \Omega,\underline{\lambda},\xi)$ & $\hspace{-0.35cm} \= \sin(\chiII)\( \df \chiII^{2}\sinh^{2}(\lambda\zetaI) - \zetaI^{2}\cosh^{2}(\lambda\zetaI)\) - \zetaI\chiII\cos(\chiII)\sinh(2\lambda\zetaI)$ \\
    ${\cal D}^{\TTM}_{\E\O\E,n_x,n_y}( \Omega,\underline{\lambda},\xi)$ & $\hspace{-0.35cm} \= \sin(\chiII)\( \df \kr^{2}\chiII^{2}\cosh^{2}(\lambda\zetaI) - \zetaI^{2}\sinh^{2}(\lambda\zetaI)\) - \kr\zetaI\chiII\cos(\chiII)\sinh(2\lambda\zetaI)$.
\end{tabular}
\end{eqnarray*}

\subsection{$\man{N}$=EEE Cavity Discrete  Spectrum Generators}
For this configuration, the systems for the TE and TM modes respectively are:
\begin{eqnarray*}
 \scriptsize
\begin{tabular}{l}
    $\( \!\!\! \begin{array}{cccccc}
            1 & 0 & -1 & 0 & 0 & 0 \\
            0 & 0 & \cosh(\zetaII) & \sinh(\zetaII) & -\cosh(\zetaI) & -\sinh(\zetaI) \\
            0 & \zetaI & 0 & -\zetaII & 0 & 0 \\
            0 & 0 & \zetaII\sinh(\zetaII) & \zetaII\cosh(\zetaII) & -\zetaI\sinh(\zetaI) & -\zetaI\cosh(\zetaI) \\
            \cosh(\lambda\zetaI) & -\sinh(\lambda\zetaI) & 0 & 0 & 0 & 0 \\
            0 & 0 & 0 & 0 & \cosh(\eta_{\E}) & \sinh(\eta_{\E}) \\
        \end{array} \!\!\! \) \!\! \( \!\!\!   \begin{array}{c}
                                \MMD{P}{\TTE}{\I}{\Omega}{\E} \\
                                \MMD{Q}{\TTE}{\I}{\Omega}{\E} \\
                                \MMD{P}{\TTE}{\II}{\Omega}{\E} \\
                                \MMD{Q}{\TTE}{\II}{\Omega}{\E} \\
                                \MMD{P}{\TTE}{\III}{\Omega}{\E} \\
                                \MMD{Q}{\TTE}{\III}{\Omega}{\E} \\
                            \end{array} \!\!\! \) \!\! \= 0$ \\
                             \\
    $\( \!\!\! \begin{array}{cccccc}
            0 & \zetaI & 0 & -\zetaII & 0 & 0 \\
            0 & 0 & \zetaII\sinh(\zetaII) & \zetaII\cosh(\zetaII) & -\zetaI\sinh(\zetaI) & -\zetaI\cosh(\zetaI) \\
            \kr & 0 & -1 & 0 & 0 & 0 \\
            0 & 0 & \cosh(\zetaII) & \sinh(\zetaII) & -\kr\cosh(\zetaI) & -\kr\sinh(\zetaI) \\
            \sinh(\lambda\zetaI) & -\cosh(\lambda\zetaI) & 0 & 0 & 0 & 0 \\
            0 & 0 & 0 & 0 & \sinh(\eta_{\E}) & \cosh(\eta_{\E}) \\
        \end{array} \!\!\! \) \!\! \( \!\!\!   \begin{array}{c}
                                \MMD{P}{\TTM}{\I}{\Omega}{\E} \\
                                \MMD{Q}{\TTM}{\I}{\Omega}{\E} \\
                                \MMD{P}{\TTM}{\II}{\Omega}{\E} \\
                                \MMD{Q}{\TTM}{\II}{\Omega}{\E} \\
                                \MMD{P}{\TTM}{\III}{\Omega}{\E} \\
                                \MMD{Q}{\TTM}{\III}{\Omega}{\E} \\
                            \end{array} \!\!\! \) \!\! \= 0$
\end{tabular}
\end{eqnarray*}
where $\eta_{\E}=(1+\lambda)\zetaI$ with corresponding determinants:
\begin{eqnarray*}
 \footnotesize
\begin{tabular}{ll}
    ${\cal D}^{\TTE}_{\E\E\E,n_x,n_y}( \Omega,\underline{\lambda},\xi)$ & $\hspace{-0.35cm} \= \sinh(\zetaII)\( \df \zetaII^{2}\sinh^{2}(\lambda\zetaI) + \zetaI^{2}\cosh^{2}(\lambda\zetaI)\) + \zetaI\zetaII\cosh(\zetaII)\sinh(2\lambda\zetaI)$ \\
    ${\cal D}^{\TTM}_{\E\E\E,n_x,n_y}( \Omega,\underline{\lambda},\xi)$ & $\hspace{-0.35cm} \= \sinh(\zetaII)\( \df \kr^{2}\zetaII^{2}\cosh^{2}(\lambda\zetaI) + \zetaI^{2}\sinh^{2}(\lambda\zetaI)\) + \kr\zetaI\zetaII\cosh(\zetaII)\sinh(2\lambda\zetaI)$.
\end{tabular}
\end{eqnarray*}
Since $\zetaI,\zetaII>0$ and $\cosh(x),\sinh(x)>0$ for all $x>0$, if the permittivity is everywhere positive, the above determinants are also positive and hence have no real zeroes for $\Omega$. Therefore, {\it with everywhere positive $\kr$}, no modes in the $\man{N}$=EEE configuration can exist within the cavity. \\
Although there can be no such modes with real spectra for everywhere positive $\kr$, if one contemplates dispersive or meta-materials (where relative permittivities $\k(\omega)$ may not be positive for all $\omega$) such spectral-modes may exist in the TM configuration.
A further simplification arises here as the size of the cavity is dilated by first letting $\Lf\to\infty$ with fixed $\Lx,\Ly,\Lz$ and then letting $\Lx,\Ly \to \infty$. In this limit, the above TM spectrum generator yields:
\begin{eqnarray*}
    (\zetaI^{2}+\kr^{2}(\Omega)\zetaII^{2})\sinh(\zetaII) + 2\kr(\Omega)\zetaI\zetaII\cosh(\zetaII) &=& 0,
\end{eqnarray*}
or
\begin{eqnarray}\label{VK}
    e^{2\zetaII} &=& \(\frac{\zetaI - \kr\zetaII}{\zetaI + \kr\zetaII}\)^{2}.
\end{eqnarray}
and in the large $\Lf,\Lx,\Ly$ limit, the spectrum becomes continuous with $k^{2}=\pi^2 \,\(\frac{\nx^2}{\Lx^2} + \frac{\ny^2}{\Ly^2}\)$ tending to $\kx^{2}+\ky^{2}$ with $\{\kx,\ky\}\in\{\real^{+}\!\times\real^{+}\}$ . The generator (\ref{VK}) simplifies further in the non-relativistic limit $\cc\rightarrow\infty$ to:
\begin{eqnarray}\label{vknonrel}
    e^{2k\Lz} &=& \(\frac{1 - \k}{1 + \k}\)^{2}
\end{eqnarray}
with $k>0$. If one models the dispersive dielectric media by a simple (single pole) Lorentz permittivity with {\it real} constant $\omega_0$ and $\k_{0} > 0,\,\k_{0} \neq 1$:
\begin{eqnarray*}\label{model}
    \k &=& 1 + \frac{\omega_0^2( \k_{0}-1)}{\omega_0^2-\omega^2}
\end{eqnarray*}
then (\ref{VK}) has two real solutions for each  $k_x,\,k_y$:
\begin{eqnarray*}
    \omega_{\pm}(k,\Lz)=\frac{\omega_0}{\sqrt{2}}\sqrt{ \k_{0}+1 \pm (\k_{0} -1) \, e^{-k\Lz} }.
\end{eqnarray*}
With such an analytic solution one can contemplate a direct evaluation of the {\it total} quantum expectation value of a renormalized energy,
associated with all such EEE electromagnetic modes  in a long perfectly conducting cavity with a rectangular cross-section of area $\Lx\Ly$, defined as:
\begin{eqnarray*}
    \man{E}(\Lz) &=& \frac{\hbar}{2} \sum_{\nx=1}^\infty \,   \sum_{\ny=1}^\infty \,\left\{ \df \[ \df\omega_{+}(k,\Lz) - \omega_{+}(k,\infty)  \,\]
         +  \[ \df\omega_{-}(k,\Lz) - \omega_{-}(k,\infty)\,\]  \right\}\\
      &=& \frac{\hbar \omega_0}{2} \sqrt{ \frac{\k_{0} +1}{2} }\sum_{\nx=1}^\infty \,   \sum_{\ny=1}^\infty \, ( \sqrt{1+\delta e^{-k\,\Lz} } +
      \sqrt{1-\delta e^{-k\,\Lz} } -2 )\\
      &=& - \hbar \omega_0 \sqrt{ \frac{\k_{0} +1}{2} }\sum_{\nx=1}^\infty \,   \sum_{\ny=1}^\infty \,\sum_{n=1}^\infty a_{2n} \, \delta^{2n}\, e^{-
      n k \Lz}
\end{eqnarray*}
where $\delta =\frac{\k_{0}-1}{\k_{0}+1}$ and $ a_n= (2n)!/(4^n (2n-1) (n!)^2)  $.
As $\Lx,\,\Ly \to \infty $ the sums $\sum_{\nx}, \sum_{\ny} $ over discrete integers tend to the integral operators
$\int_0^\infty \frac{\Lx}{\pi} \,d\kx, \,\,\int_0^\infty \frac{\Ly}{\pi}\, d\ky, \,$ respectively and the total energy per unit area,
$\frac{\man{E}(\Lz)}{\Lx \Ly }$,  tends to $\man{U}(\Lz)$ where
\begin{eqnarray*}
    \man{U}(\Lz) &=& -\frac{\hbar\omega_0}{8\pi \Lz^2 } \sqrt{\frac{\k_{0}+1}{2}  }\sum_{n=1}^\infty \frac{a_{2n}}{n^2}\, \delta^{2n}
\end{eqnarray*}
yielding a finite stress of magnitude $\left| \frac{\partial \man{U}}{\partial \Lz}|_{\tm{\Sigma}} \right| $ across each medium discontinuity at $\Sigma$.

This is precisely the non-relativistic (non-retarded) {\it plasmon} induced energy density found by Van Kampen \cite{Kampen} (albeit derived from Maxwell modes in a large perfectly conducting cavity).

\subsection{$\man{N}$=ELE Cavity Discrete  Spectrum Generators}
In this configuration, $\chiII=0$ (i.e. $\Omega=R$) and the systems for the TE and TM modes respectively are:
\begin{eqnarray*}
 \small
\begin{tabular}{l}
    $\( \!\!\! \begin{array}{cccccc}
            1 & 0 & -1 & 0 & 0 & 0 \\
            0 & 0 & 1 & 1 & -\cosh(\zetaI) & -\sinh(\zetaI) \\
            0 & \zetaI & 0 & -1 & 0 & 0 \\
            0 & 0 & 0 & 1 & -\zetaI\sinh(\zetaI) & -\zetaI\cosh(\zetaI) \\
            \cosh(\lambda\zetaI) & -\sinh(\lambda\zetaI) & 0 & 0 & 0 & 0 \\
            0 & 0 & 0 & 0 & \cosh(\eta_{\E}) & \sinh(\eta_{\E}) \\
        \end{array} \!\!\!\) \!\! \( \!\!\!  \begin{array}{c}
                                \MMD{P}{\TTE}{\I}{\Omega}{\E} \\
                                \MMD{Q}{\TTE}{\I}{\Omega}{\E} \\
                                \MMD{P}{\TTE}{\II}{\Omega}{\L} \\
                                \MMD{Q}{\TTE}{\II}{\Omega}{\L} \\
                                \MMD{P}{\TTE}{\III}{\Omega}{\E} \\
                                \MMD{Q}{\TTE}{\III}{\Omega}{\E} \\
                            \end{array} \!\!\! \)  \= 0$ \\
                              \\
    $\( \!\!\! \begin{array}{cccccc}
            0 & \zetaI & 0 & -1 & 0 & 0 \\
            0 & 0 &  & 1 & -\zetaI\sinh(\zetaI) & -\zetaI\cosh(\zetaI) \\
            \kr & 0 & -1 & 0 & 0 & 0 \\
            0 & 0 & 1 & 1 & -\kr\cosh(\zetaI) & -\kr\sinh(\zetaI) \\
            \sinh(\lambda\zetaI) & -\cosh(\lambda\zetaI) & 0 & 0 & 0 & 0 \\
            0 & 0 & 0 & 0 & \sinh(\eta_{\E}) & \cosh(\eta_{\E}) \\
        \end{array} \!\!\! \) \!\! \( \!\!\!  \begin{array}{c}
                                \MMD{P}{\TTM}{\I}{\Omega}{\E} \\
                                \MMD{Q}{\TTM}{\I}{\Omega}{\E} \\
                                \MMD{P}{\TTM}{\II}{\Omega}{\L} \\
                                \MMD{Q}{\TTM}{\II}{\Omega}{\L} \\
                                \MMD{P}{\TTM}{\III}{\Omega}{\E} \\
                                \MMD{Q}{\TTM}{\III}{\Omega}{\E} \\
                            \end{array} \!\!\!\)  \= 0$
\end{tabular}
\end{eqnarray*}
where $\eta_{\E}=(1+\lambda)\zetaI$ with corresponding determinants:
\begin{eqnarray*}
 \small
\begin{tabular}{ll}
    ${\cal D}^{\TTE}_{\E\L\E,n_x,n_y}( \Omega,\underline{\lambda},\xi)$ & $\hspace{-0.35cm} \= \zetaI\cosh(\lambda\zetaI)\( \df 2\sinh(\lambda\zetaI) + \zetaI\cosh(\lambda\zetaI) \) $  \\
    ${\cal D}^{\TTE}_{\E\L\E,n_x,n_y}( \Omega,\underline{\lambda},\xi)$ & $\hspace{-0.35cm} \= \zetaI\sinh(\lambda\zetaI)\( \df \zetaI\sinh(\lambda\zetaI) + 2\kr\cosh(\lambda\zetaI) \)$.
\end{tabular}
\end{eqnarray*}
Since both $\sinh(\lambda\zetaI),\cosh(\lambda\zetaI)>0$ for all $\zetaI>0$ (since $\lambda>0$), these spectrum generating functions have no real zeroes for either branch if $\kr >0$. In particular, no non-trivial modes in the $\man{N}$=ELE configuration can exist within a cavity in the constant dielectric model with positive permittivity. As before, non-trivial TM solutions may arise when $\kr(\Omega)$ is not positive definite.

\subsection{$\man{N}$=LOL Cavity Discrete  Spectrum Generators}
In this configuration, $\chiI=\zetaI=0$ (i.e. $\Omega=\frac{R}{\sqrt{\kr\,}}$) and the systems for the TE and TM modes respectively are
\begin{eqnarray*}
 \small
\begin{tabular}{l}
    $\( \!\!\! \begin{array}{cccccc}\footnotesize
            1 & 0 & -1 & 0 & 0 & 0 \\
            0 & 0 & \cos(\chiII) & \sin(\chiII) & -1 & -1 \\
            0 & 1 & 0 & -\chiII & 0 & 0 \\
            0 & 0 & \chiII\sin(\chiII) & -\chiII\cos(\chiII) & 0 & 1 \\
            1 & -\lambda & 0 & 0 & 0 & 0 \\
            0 & 0 & 0 & 0 & 1 & 1+\lambda \\
        \end{array} \!\!\!\) \!\!\( \!\!\!  \begin{array}{c}
                                \MMD{P}{\TTE}{\I}{\Omega}{\L} \\
                                \MMD{Q}{\TTE}{\I}{\Omega}{\L} \\
                                \MMD{P}{\TTE}{\II}{\Omega}{\O} \\
                                \MMD{Q}{\TTE}{\II}{\Omega}{\O} \\
                                \MMD{P}{\TTE}{\III}{\Omega}{\L} \\
                                \MMD{Q}{\TTE}{\III}{\Omega}{\L} \\
                            \end{array} \!\!\!\) \= 0$ \\
                             \\
   $\( \!\!\! \begin{array}{cccccc}\footnotesize
            0 & 1 & 0 & -\chiII & 0 & 0 \\
            0 & 0 & \chiII\sin(\chiII) & -\chiII\cos(\chiII) & 0 & 1 \\
            \kr & 0 & -1 & 0 & 0 & 0 \\
            0 & 0 & \cos(\chiII) & \sin(\chiII) & -\kr & -\kr \\
            0 & 1 & 0 & 0 & 0 & 0 \\
            0 & 0 & 0 & 0 & 0 & 1 \\
        \end{array} \!\!\!\)\!\! \(  \!\!\! \begin{array}{c}
                                \MMD{P}{\TTM}{\I}{\Omega}{\L} \\
                                \MMD{Q}{\TTM}{\I}{\Omega}{\L} \\
                                \MMD{P}{\TTM}{\II}{\Omega}{\O} \\
                                \MMD{Q}{\TTM}{\II}{\Omega}{\O} \\
                                \MMD{P}{\TTM}{\III}{\Omega}{\L} \\
                                \MMD{Q}{\TTM}{\III}{\Omega}{\L} \\
                            \end{array} \!\!\! \) \= 0$
\end{tabular}
\end{eqnarray*}
with corresponding determinants:
\begin{eqnarray*}
\small
\begin{tabular}{ll}
    ${\cal D}^{\TTE}_{\L\O\L,n_x,n_y}( \Omega,\underline{\lambda},\xi)$ & $ \hspace{-0.35cm} \= \sin(\chiII)\( \df \lambda^2\chiII^{2}-1 \) - 2\lambda\chiII\cos(\chiII)$ \\
    ${\cal D}^{\TTE}_{\L\O\L,n_x,n_y}( \Omega,\underline{\lambda},\xi)$ & $ \hspace{-0.35cm} \= \sin(\chiII)$.
\end{tabular}
\end{eqnarray*}
As in the $\man{N}=\text{OLO}$ case, the vanishing of the determinants here yields a class of Diophantine type equations. For example, in the constant dielectric model with $\kr=\xi$, solutions to the TM equation are found by requiring that $\sin(\chiII)=\sin\(R\sqrt{\frac{1}{\xi}-1}\)=0$, that is $R\sqrt{\frac{1}{\xi}-1} = \ell\pi$
for some integer $\ell\geq 0$. From (\ref{rRdef}), this yields
\begin{eqnarray*}
    \nx^{2} + \lambda_{2}^{2}\ny^{2} \= \lambda_{1}^{2}\ell^{2}\(\frac{1}{\xi}-1\)^{-1},
\end{eqnarray*}
which is a Diophantine type equation in terms of the three integers $(\nx,\ny,\ell)$, the constant relative permittivity $\xi$ and the geometric factors $\lambda_{1},\lambda_{2}$. For a particular choice of $\xi,\,\lambda_{1},\,\lambda_{2}$, there may or may not be any integer solutions yielding a real $\Omega=\frac{l\,\pi}{\sqrt{1-\xi }}$. For example, with $\xi=\frac{1}{2},\lambda_{2}=1$, one has
\begin{eqnarray*}
    \nx^{2} + \ny^{2} \= \lambda_{1}^{2}\ell^{2}.
\end{eqnarray*}
For $\lambda_{1}=1$ this is a Pythagorean equation with many integer solutions, but for a general $\lambda_{1}$, this is not the case.

\subsection{$\man{N}$=LEL Cavity Discrete  Spectrum Generators}
In this configuration, $\chiI=\zetaI=0$ (i.e. $\Omega=\frac{R}{\sqrt{\kr\,}}$) and the systems for the TE and TM modes respectively are:
\begin{eqnarray*}
 \small
\begin{tabular}{l}
    $\( \!\!\! \begin{array}{cccccc}
            1 & 0 & -1 & 0 & 0 & 0 \\
            0 & 0 & \cosh(\zetaII) & \sinh(\zetaII) & -1 & -1 \\
            0 & 1 & 0 & -\zetaII & 0 & 0 \\
            0 & 0 & \zetaII\sinh(\zetaII) & \zetaII\cosh(\zetaII) & 0 & -1 \\
            1 & -\lambda & 0 & 0 & 0 & 0 \\
            0 & 0 & 0 & 0 & 1 & 1+\lambda \\
        \end{array} \!\!\! \)\!\! \( \!\!\!  \begin{array}{c}
                                \MMD{P}{\TTE}{\I}{\Omega}{\L} \\
                                \MMD{Q}{\TTE}{\I}{\Omega}{\L} \\
                                \MMD{P}{\TTE}{\II}{\Omega}{\E} \\
                                \MMD{Q}{\TTE}{\II}{\Omega}{\E} \\
                                \MMD{P}{\TTE}{\III}{\Omega}{\L} \\
                                \MMD{Q}{\TTE}{\III}{\Omega}{\L} \\
                            \end{array} \!\!\!\) \= 0$ \\
                            \\
    $\(\!\!\!  \begin{array}{cccccc}
            0 & 1 & 0 & -\zetaII & 0 & 0 \\
            0 & 0 & \zetaII\sinh(\zetaII) & \zetaII\cosh(\zetaII) & 0 & -1 \\
            \kr & 0 & -1 & 0 & 0 & 0 \\
            0 & 0 & \cosh(\zetaII) & \sinh(\zetaII) & -\kr & -\kr \\
            0 & 1 & 0 & 0 & 0 & 0 \\
            0 & 0 & 0 & 0 & 0 & 1 \\
        \end{array} \!\!\!\) \!\!\( \!\!\!  \begin{array}{c}
                                \MMD{P}{\TTM}{\I}{\Omega}{\L} \\
                                \MMD{Q}{\TTM}{\I}{\Omega}{\L} \\
                                \MMD{P}{\TTM}{\II}{\Omega}{\E} \\
                                \MMD{Q}{\TTM}{\II}{\Omega}{\E} \\
                                \MMD{P}{\TTM}{\III}{\Omega}{\L} \\
                                \MMD{Q}{\TTM}{\III}{\Omega}{\L} \\
                            \end{array} \!\!\! \) \= 0$
\end{tabular}
\end{eqnarray*}
with corresponding determinants:
\begin{eqnarray*}
\small
\begin{tabular}{ll}
    ${\cal D}^{\TTE}_{\L\E\L,n_x,n_y}( \Omega,\underline{\lambda},\xi)$ & $\hspace{-0.35cm} \= \sinh(\zetaII)\(  \lambda^2\zetaII^{2}+1 \df \) + 2\lambda\zetaII\cosh(\zetaII)$ \\
    ${\cal D}^{\TTM}_{\L\E\L,n_x,n_y}( \Omega,\underline{\lambda},\xi)$ & $\hspace{-0.35cm} \= \sinh(\zetaII)$.
\end{tabular}
\end{eqnarray*}
Since $\lambda>0$ and $\sinh(\zetaII),\cosh(\zetaII)>0$ for all $\zetaII>0$, these spectrum generating functions have no real zeroes for any $\kr$. Thus, no non-trivial modes in the $\man{N}$=LEL configuration can exist within the cavity.

\newpage
\section{Modes in Open Domains}\label{OpenAppendix}
Following (\ref{chizetadef}) and the discussion in section~\ref{CavitySect}, the configuration of dielectric half-spaces has  $\kr^{\I}(\Omega,\xi)=\kr^{\III}(\Omega,\xi)\equiv \kr(\Omega,\xi)$ and $\kr^{\II}=1$ with $\kr$ dependent upon a single material constant $\xi$.

\subsection{Implicit Equations for the Local Dispersion Relations with Category P modes}
\subsubsection{$\man{N}$=OOO}
With
\begin{eqnarray*}
    \MMD{\hpsi}{s}{\II}{\Omega}{\O}{(Z)} &\= \MMD{P}{s}{\II}{\Omega}{\O}\, e^{-i\chiII Z} + \MMD{Q}{s}{\II}{\Omega}{\O}\, e^{i\chiII Z},
\end{eqnarray*}
the four interface conditions (\ref{IFCond}) yield the inhomogeneous system of equations for the TE and TM modes respectively:
\begin{eqnarray*}
 \footnotesize
\begin{tabular}{ll}
    $\( \!\! \begin{array}{cccccc}
            1 & -1 & -1 & 0 \\
            0 & e^{-i\chiII} & e^{i\chiII} & -e^{i\chiI} \\
            \chiI & -\chiII & \chiII & 0 \\
            0 & \chiII e^{-i\chiII} & -\chiII e^{i\chiII} & \chiI e^{i\chiI} \\
        \end{array} \!\! \) \!\! \( \!\!  \begin{array}{c}
                                \MMD{P}{\TTE}{\I}{\Omega}{\O} \\
                                \MMD{P}{\TTE}{\II}{\Omega}{\O} \\
                                \MMD{Q}{\TTE}{\II}{\Omega}{\O} \\
                                \MMD{Q}{\TTE}{\III}{\Omega}{\O} \\
                            \end{array} \!\! \)$ & $\hspace{-0.35cm} \=
                                               \( \!\!  \begin{array}{c}
                                                        -\MMD{Q}{\TTE}{\I}{\Omega}{\O} \\
                                                        e^{-i\chiI}\MMD{P}{\TTE}{\III}{\Omega}{\O} \\
                                                        \chiI\MMD{Q}{\TTE}{\I}{\Omega}{\O} \\
                                                        \chiI e^{-i\chiI}\MMD{P}{\TTE}{\III}{\Omega}{\O} \\
                                                    \end{array} \!\! \)$ \\
                            & \\
    $\( \!\! \begin{array}{cccccc}
            \chiI & -\chiII & \chiII & 0 \\
            0 & \chiII e^{i\chiII} & -\chiII e^{-i\chiII} & \chiI e^{i\chiI} \\
            \kr & -1 & -1 & 0 \\
            0 & e^{-i\chiII} & e^{i\chiII} & -\kr e^{i\chiI} \\
        \end{array} \!\!\) \!\! \( \!\!  \begin{array}{c}
                                \MMD{P}{\TTM}{\I}{\Omega}{\O} \\
                                \MMD{P}{\TTM}{\II}{\Omega}{\O} \\
                                \MMD{Q}{\TTM}{\II}{\Omega}{\O} \\
                                \MMD{Q}{\TTM}{\III}{\Omega}{\O} \\
                            \end{array} \!\!\)$ & $\hspace{-0.35cm} \=
                                               \(  \!\! \begin{array}{c}
                                                        \chiI\MMD{Q}{\TTM}{\I}{\Omega}{\O} \\
                                                        \chiI e^{-i\chiI}\MMD{P}{\TTM}{\III}{\Omega}{\O} \\
                                                        -\kr\MMD{Q}{\TTM}{\I}{\Omega}{\O} \\
                                                        \kr e^{-i\chiI}\MMD{P}{\TTM}{\III}{\Omega}{\O} \\
                                                    \end{array}\!\! \)$
\end{tabular}
\end{eqnarray*}
with solutions for $\MMD{P}{s}{\I}{\Omega}{\O},\MMD{P}{s}{\II}{\Omega}{\O},\MMD{Q}{s}{\II}{\Omega}{\O},\MMD{Q}{s}{\III}{\Omega}{\O}$ in terms of $\MMD{Q}{s}{\I}{\Omega}{\O},\MMD{P}{s}{\III}{\Omega}{\O}$. Using the substitutions (\ref{P_RL}) yields the reflection and transmission coefficients:
\begin{eqnarray*}
 \scriptsize
\begin{tabular}{llll}
    $R^{\TTE,u}_{\O\O\O}$ & $\displaystyle \hspace{-0.4cm} \= \frac{2i\sin(\chiII)(\chiII^{2}-\chiI^{2})}{ (\chiI+\chiII)^{2}e^{-i\chiII} - (\chiI-\chiII)^{2}e^{i\chiII}}$, & $\hspace{-0.15cm} T^{\TTE,u}_{\O\O\O}$ & $\displaystyle \hspace{-0.4cm} \= \frac{4\chiI\chiII e^{-i\chiI}}{ (\chiI+\chiII)^{2}e^{-i\chiII} - (\chiI-\chiII)^{2}e^{i\chiII} }$, \\
    & & & \\
    $R^{\TTM,u}_{\O\O\O}$ & $\displaystyle \hspace{-0.4cm} \= \frac{2i\sin(\chiII)(\kr^{2}\chiII^{2}-\chiI^{2})}{ (\chiI+\kr\chiII)^{2}e^{-i\chiII} -
     (\chiI-\kr\chiII)^{2}e^{i\chiII}}$, & $\hspace{-0.15cm} T^{\TTM,u}_{\O\O\O}$ & $\displaystyle \hspace{-0.4cm} \= \frac{4\kr\chiI\chiII e^{-i\chiI}}{ (\chiI+\kr\chiII)^{2}e^{-i\chiII} - (\chiI-\kr\chiII)^{2}e^{i\chiII} }$,
\end{tabular}
\end{eqnarray*}
with $u\in\{\text{R,L}\}$ denoting either R or L propagating waves.

\subsubsection{$\man{N}$=OEO}
With
\begin{eqnarray*}
    \MMD{\hpsi}{s}{\II}{\Omega}{\E}{(Z)} &\= \MMD{P}{s}{\II}{\Omega}{\E}\, e^{-\zetaII Z} + \MMD{Q}{s}{\II}{\Omega}{\E}\, e^{\zetaII Z},
\end{eqnarray*}
the four interface conditions (\ref{IFCond}) yield the inhomogeneous system of equations for the TE and TM modes respectively:
\begin{eqnarray*}
 \footnotesize
\begin{tabular}{ll}
    $\( \!\! \begin{array}{cccccc}
            1 & -1 & -1 & 0 \\
            0 & e^{-\zetaII} & e^{-\zetaII} & -e^{i\chiI} \\
            \chiI & i\zetaII & -i\zetaII & 0 \\
            0 & i\zetaII e^{-\zetaII} & -i\zetaII e^{\zetaII} & -\chiI e^{i\chiI} \\
        \end{array} \!\!\)\!\! \( \!\!  \begin{array}{c}
                                \MMD{P}{\TTE}{\I}{\Omega}{\O} \\
                                \MMD{P}{\TTE}{\II}{\Omega}{\E} \\
                                \MMD{Q}{\TTE}{\II}{\Omega}{\E} \\
                                \MMD{Q}{\TTE}{\III}{\Omega}{\O} \\
                            \end{array} \!\!\)$ & $\hspace{-0.35cm} \=
                                               \(  \!\! \begin{array}{c}
                                                        -\MMD{Q}{\TTE}{\I}{\Omega}{\O} \\
                                                        e^{-i\chiI}\MMD{P}{\TTE}{\III}{\Omega}{\O} \\
                                                        \chiI\MMD{Q}{\TTE}{\I}{\Omega}{\O} \\
                                                        -\chiI e^{-i\chiI}\MMD{P}{\TTE}{\III}{\Omega}{\O} \\
                                                    \end{array}\!\! \)$ \\
                            & \\
    $\( \!\! \begin{array}{cccccc}
            \chiI & i\zetaII & -i\zetaII & 0 \\
            0 & i\zetaII e^{-\zetaII} & -i\zetaII e^{\zetaII} & -\chiI e^{i\chiI} \\
            \kr & -1 & -1 & 0 \\
            0 & e^{-\zetaII} & e^{\zetaII} & -\kr^{\I} e^{i\chiI} \\
        \end{array} \!\!\) \!\!\( \!\!  \begin{array}{c}
                                \MMD{P}{\TTM}{\I}{\Omega}{\O} \\
                                \MMD{P}{\TTM}{\II}{\Omega}{\E} \\
                                \MMD{Q}{\TTM}{\II}{\Omega}{\E} \\
                                \MMD{Q}{\TTM}{\III}{\Omega}{\O} \\
                            \end{array}\!\! \)$ & $\hspace{-0.35cm} \=
                                               \(  \!\! \begin{array}{c}
                                                        \chiI\MMD{Q}{\TTM}{\I}{\Omega}{\O} \\
                                                        -i\chiI e^{-i\chiI}\MMD{P}{\TTM}{\III}{\Omega}{\O} \\
                                                        -\kr\MMD{Q}{\TTM}{\I}{\Omega}{\O} \\
                                                        \kr e^{-i\chiI}\MMD{P}{\TTM}{\III}{\Omega}{\O} \\
                                                    \end{array} \!\! \)$
\end{tabular}
\end{eqnarray*}
with solutions for $\MMD{P}{s}{\I}{\Omega}{\O},\MMD{P}{s}{\II}{\Omega}{\E},\MMD{Q}{s}{\II}{\Omega}{\E},\MMD{Q}{s}{\III}{\Omega}{\O}$ in terms of $\MMD{Q}{s}{\I}{\Omega}{\O},\MMD{P}{s}{\III}{\Omega}{\O}$. Using the substitutions (\ref{P_RL}) yields the reflection and transmission coefficients:
\begin{eqnarray*}
 \scriptsize
\begin{tabular}{llll}
    $R^{\TTE,u}_{\O\E\O}$ & $\displaystyle \hspace{-0.4cm} \= -\frac{2(\chiI^{2}+\zetaII^{2})\sinh(\zetaII)}{ (\chiI + i\zetaII)^{2} e^{\zetaII} - (\chiI - i\zetaII)^{2} e^{-\zetaII}}$,  & $T^{\TTE,u}_{\O\E\O}$ & $\displaystyle \hspace{-0.4cm} \= \frac{4i\chiI\zetaII e^{-i\chiI}}{ (\chiI + i\zetaII)^{2} e^{\zetaII} - (\chiI - i\zetaII)^{2} e^{-\zetaII} }$ \\
    & & & \\
    $R^{\TTE,u}_{\O\E\O}$ & $\displaystyle \hspace{-0.4cm} \= -\frac{2(\chiI^{2}+\kr^{2}\zetaII^{2})\sinh(\zetaII)}{ (\chiI + i\kr\zetaII)^{2} e^{\zetaII} - (\chiI - i\kr\zetaII)^{2}e^{-\zetaII} }$, & $T^{\TTE,u}_{\O\E\O}$ & $\displaystyle \hspace{-0.4cm} \= \frac{4i\kr\chiI\zetaII e^{-i\chiI}}{ (\chiI + i\kr\zetaII)^{2} e^{\zetaII} - (\chiI - i\kr\zetaII)^{2} e^{-\zetaII}  }$,
\end{tabular}
\end{eqnarray*}
with $u\in\{\text{R,L}\}$ denoting either R or L propagating waves.

\subsubsection{$\man{N}$=OLO}
With $\zetaII=\chiII=0$ (i.e. $\Omega=R$) and
\begin{eqnarray*}
    \MMD{\hpsi}{s}{\II}{\Omega}{\L}{(Z)} &\= \MMD{P}{s}{\II}{\Omega}{\L} + \MMD{Q}{s}{\II}{\Omega}{\L}\, Z,
\end{eqnarray*}
the four interface conditions (\ref{IFCond}) yield the inhomogeneous system of equations for the TE and TM modes respectively:
\begin{eqnarray*}
  \footnotesize
\begin{tabular}{ll}
    $\( \!\! \begin{array}{cccccc}
            1 & -1 & 0 & 0 \\
            0 & 1 & 1 & -e^{i\chiI} \\
            \chiI & 0 & -i & 0 \\
            0 & 0 & i & \chiI e^{i\chiI} \\
        \end{array} \!\!\)\!\! \( \!\!  \begin{array}{c}
                                \MMD{P}{\TTE}{\I}{\Omega}{\O} \\
                                \MMD{P}{\TTE}{\II}{\Omega}{\L} \\
                                \MMD{Q}{\TTE}{\II}{\Omega}{\L} \\
                                \MMD{Q}{\TTE}{\III}{\Omega}{\O} \\
                            \end{array}\!\! \)$ & $\hspace{-0.35cm} \=
                                               \(  \!\! \begin{array}{c}
                                                        -\MMD{Q}{\TTE}{\I}{\Omega}{\O} \\
                                                        e^{-i\chiI}\MMD{P}{\TTE}{\III}{\Omega}{\O} \\
                                                        \chiI\MMD{Q}{\TTE}{\I}{\Omega}{\O} \\
                                                        \chiI e^{-i\chiI}\MMD{P}{\TTE}{\III}{\Omega}{\O} \\
                                                    \end{array}\!\! \)$ \\
                            & \\
    $\( \!\! \begin{array}{cccccc}
            \chiI & 0 & -i & 0 \\
            0 & 0 & i & \chiI e^{i\chiI} \\
            \kr & -1 & 0 & 0 \\
            0 & 1 & 1 & -\kr e^{i\chiI} \\
        \end{array} \!\!\) \!\!\(\!\!   \begin{array}{c}
                                \MMD{P}{\TTM}{\I}{\Omega}{\O} \\
                                \MMD{P}{\TTM}{\II}{\Omega}{\L} \\
                                \MMD{Q}{\TTM}{\II}{\Omega}{\L} \\
                                \MMD{Q}{\TTM}{\III}{\Omega}{\O} \\
                            \end{array} \!\!\)$ & $\hspace{-0.35cm} \=
                                               \( \!\!  \begin{array}{c}
                                                        \chiI\MMD{Q}{\TTM}{\I}{\Omega}{\O} \\
                                                        \chiI e^{-i\chiI}\MMD{P}{\TTM}{\III}{\Omega}{\O} \\
                                                        -\kr\MMD{Q}{\TTM}{\I}{\Omega}{\O} \\
                                                        \kr e^{-i\chiI}\MMD{P}{\TTM}{\III}{\Omega}{\O} \\
                                                    \end{array}\!\! \)$
\end{tabular}
\end{eqnarray*}
with solutions for $\MMD{P}{s}{\I}{\Omega}{\O},\MMD{P}{s}{\II}{\Omega}{\L},\MMD{Q}{s}{\II}{\Omega}{\L},\MMD{Q}{s}{\III}{\Omega}{\O}$ in terms of $\MMD{Q}{s}{\I}{\Omega}{\O},\MMD{P}{s}{\III}{\Omega}{\O}$. Using the substitutions (\ref{P_RL}) yields the reflection and transmission coefficients:
\begin{eqnarray*}
 \footnotesize
\begin{tabular}{llll}
    $R^{\TTE,u}_{\O\L\O}$ & $\displaystyle \hspace{-0.35cm}\= \frac{i\chiI }{ i\chiI - 2 }$, & $\; T^{\TTE,u}_{\O\L\O}$ & $\displaystyle\hspace{-0.35cm}\= -\frac{2 e^{-i\chiI}}{ i\chiI - 2 }$ \\
    & & & \\
    $R^{\TTM,u}_{\O\L\O}$ & $\displaystyle\hspace{-0.35cm}\= \frac{i\chiI }{ i\chiI - 2\kr }$, & $\; T^{\TTM,u}_{\O\L\O}$ & $\displaystyle\hspace{-0.35cm}\= -\frac{2\kr e^{-i\chiI}}{ i\chiI - 2\kr }$ ,
\end{tabular}
\end{eqnarray*}
with $u\in\{\text{R,L}\}$ denoting either R or L propagating waves.

\subsection{Implicit Equations for the Global Dispersion Relations $\omega^{s}_{\man{N}}=\wt{\omega}^{s}_{\man{N}}(\kx,\ky)$ with Category E modes}
\subsubsection{$\man{N}$=EOE}
With
\begin{eqnarray*}
    \MMD{\hpsi}{s}{\II}{\Omega}{\O}{(Z)} &\= \MMD{P}{s}{\II}{\Omega}{\O}\, e^{-i\chiII Z} + \MMD{Q}{s}{\II}{\Omega}{\O}\, e^{i\chiII Z},
\end{eqnarray*}
the four interface conditions (\ref{IFCond}) yield the homogeneous system of equations for the TE and TM modes respectively:
\begin{eqnarray*}
 \footnotesize
\begin{tabular}{l}
    $\( \!\! \begin{array}{cccc}
            1 & -1 & -1 & 0  \\
            0 & e^{-i\chiII} & e^{i\chiII} & -e^{-\zetaI} \\
            i\zetaI & -\chiII & \chiII & 0 \\
            0 & \chiII e^{-i\chiII} & -\chiII e^{i\chiII} & -i\zetaI e^{-\zetaI} \\
        \end{array} \!\!\) \!\!\( \!\!  \begin{array}{c}
                                \MMD{Q}{\TTE}{\I}{\Omega}{\E} \\
                                \MMD{P}{\TTE}{\II}{\Omega}{\O} \\
                                \MMD{Q}{\TTE}{\II}{\Omega}{\O} \\
                                \MMD{P}{\TTE}{\III}{\Omega}{\E} \\
                            \end{array}\!\! \) \= 0 $  \\
                             \\
   $\( \!\! \begin{array}{cccccc}
            i\zetaI & -\chiII & \chiII & 0 \\
            0 & \chiII e^{-i\chiII} & -\chiII e^{i\chiII} & i\zetaI e^{-\zetaI} \\
            \kr & -1 & -1 & 0 \\
            0 & e^{-i\chiII} & e^{i\chiII} & -\kr e^{-\zetaI} \\
        \end{array}\!\! \) \!\!\(  \!\! \begin{array}{c}
                                \MMD{Q}{\TTM}{\I}{\Omega}{\E} \\
                                \MMD{P}{\TTM}{\II}{\Omega}{\O} \\
                                \MMD{Q}{\TTM}{\II}{\Omega}{\O} \\
                                \MMD{P}{\TTM}{\III}{\Omega}{\E} \\
                            \end{array}\!\! \) \= 0$
\end{tabular}
\end{eqnarray*}
producing the global dispersion relations ${D}^{s}_{\E\O\E,\kx,\ky}( \Omega,\xi)=0$ where
\begin{eqnarray}
    \label{DTE_EOE} &&\mbox{\small $\displaystyle {D}^{\TTE}_{\E\O\E,\kx,\ky}( \Omega,\xi) \= e^{-\zetaI}\( \df (\zetaI+i\chiII)^{2}e^{i\chiII} - (\zetaI-i\chiII)^{2}e^{-i\chiII} \)$} \\
    \nonumber && \\
    \label{DTM_EOE} && \mbox{\small $\displaystyle {D}^{\TTM}_{\E\O\E,\kx,\ky}( \Omega,\xi) \= e^{-\zetaI}\( \df (\zetaI+i\kr\chiII)^{2}e^{i\chiII} - (\zetaI-i\kr\chiII)^{2}e^{-i\chiII} \)$},
\end{eqnarray}
describing allowed modes.

\subsubsection{$\man{N}$=EEE}
With
\begin{eqnarray*}
    \MMD{\hpsi}{s}{\II}{\Omega}{\E}{(Z)} &\= \MMD{P}{s}{\II}{\Omega}{\E}\,  e^{-\zetaII Z} + \MMD{Q}{s}{\II}{\Omega}{\E}\,  e^{\zetaII Z},
\end{eqnarray*}
the four interface conditions (\ref{IFCond}) yield the homogeneous system of equations for the TE and TM modes respectively:
\begin{eqnarray*}
 \footnotesize
\begin{tabular}{l}
    $\( \!\! \begin{array}{cccc}
            1 & -1 & -1 & 0  \\
            0 & e^{-\zetaII} & e^{-\zetaII} & -e^{-\zetaI} \\
            \zetaI & \zetaII & -\zetaII & 0 \\
            0 & \zetaII e^{-\zetaII} & -\zetaII e^{\zetaII} & -\zetaI e^{-\zetaI} \\
        \end{array} \!\!\) \!\!\(  \!\! \begin{array}{c}
                                \MMD{Q}{\TTE}{\I}{\Omega}{\E} \\
                                \MMD{P}{\TTE}{\II}{\Omega}{\E} \\
                                \MMD{Q}{\TTE}{\II}{\Omega}{\E} \\
                                \MMD{P}{\TTE}{\III}{\Omega}{\E} \\
                            \end{array}\!\! \) \= 0$ \\
                             \\
    $\( \!\! \begin{array}{cccccc}
            \zetaI & \zetaII & -\zetaII & 0 \\
            0 & \zetaII e^{-\zetaII} & -\zetaII e^{\zetaII} & -\zetaI e^{-\zetaI} \\
            \kr & -1 & -1 & 0 \\
            0 & e^{-\zetaII} & e^{\zetaII} & -\kr e^{-\zetaI
            } \\
        \end{array} \!\!\) \!\!\(  \!\! \begin{array}{c}
                                \MMD{Q}{\TTM}{\I}{\Omega}{\E} \\
                                \MMD{P}{\TTM}{\II}{\Omega}{\E} \\
                                \MMD{Q}{\TTM}{\II}{\Omega}{\E} \\
                                \MMD{P}{\TTM}{\III}{\Omega}{\E} \\
                            \end{array} \!\!\) \= 0$
\end{tabular}
\end{eqnarray*}
yielding the conditions for non-trivial modes ${D}^{s}_{\E\E\E,\kx,\ky}( \Omega,\xi)=0$ where :
\begin{eqnarray}
   \nonumber  {D}^{\TTE}_{\E\E\E,\kx,\ky}( \Omega,\xi) &=& e^{-\zetaI}\( \df (\zetaI+\zetaII)^{2}e^{\zetaII} - (\zetaI-\zetaII)^{2}e^{-\zetaII} \) \\
   \nonumber && \\
   \label{DTM_EEE} {D}^{\TTM}_{\E\E\E,\kx,\ky}( \Omega,\xi) &=& e^{-\zetaI}\( \df (\zetaI+\kr\zetaII)^{2}e^{\zetaII} - (\zetaI-\kr\zetaII)^{2}e^{-\zetaII} \).
\end{eqnarray}
However, since $\zetaI,\zetaII>0$, it is clear that ${D}^{\TTE}_{\E\E\E,\kx,\ky}( \Omega,\xi)$ cannot have real zeroes. The same is true for ${D}^{\TTM}_{\E\E\E,\kx,\ky}( \Omega,\xi)$ for a positive definite $\kr$. Thus, there are no $\man{N}$=EEE open configurations in such cases.

As in the bounded case, it is of interest to explore the TM modes when permittivities may be non-positive definite. The TM determinant then vanishes when
\begin{eqnarray*}
    e^{2\zetaII} &=& \(\frac{\zetaI - \kr\zetaII}{\zetaI + \kr\zetaII}\)^{2}.
\end{eqnarray*}
In the limit as $\cc\rightarrow\infty$, this becomes
\begin{eqnarray*}
    e^{2k\Lz} &=& \(\frac{1 - \k(\omega)}{1 + \k(\omega)}\)^{2},
\end{eqnarray*}
consistent with (\ref{vknonrel}).

\subsubsection{$\man{N}$=ELE}
With $\zetaII=\chiII=0$ (i.e. $\Omega=R$) and
\begin{eqnarray*}
    \MMD{\hpsi}{s}{\II}{\Omega}{\O}{(Z)} &\= \MMD{P}{s}{\II}{\Omega}{\L} + \MMD{Q}{s}{\II}{\Omega}{\L}\,Z,
\end{eqnarray*}
the four interface conditions (\ref{IFCond}) yield the homogeneous system of equations for the TE and TM modes respectively:
\begin{eqnarray*}
 \footnotesize
\begin{tabular}{l}
    $\( \!\! \begin{array}{cccc}
            1 & -1 & 0 & 0  \\
            0 & 1 & 1 & -e^{-\zetaI} \\
            \zetaI & 0 & -1 & 0 \\
            0 & 0 & 1 & \zetaI e^{-\zetaI} \\
        \end{array}\!\! \) \!\!\( \!\!  \begin{array}{c}
                                \MMD{Q}{\TTE}{\I}{\Omega}{\E} \\
                                \MMD{P}{\TTE}{\II}{\Omega}{\L} \\
                                \MMD{Q}{\TTE}{\II}{\Omega}{\L} \\
                                \MMD{P}{\TTE}{\III}{\Omega}{\E} \\
                            \end{array}\!\! \) \= 0$ \\
                             \\
    $\( \!\! \begin{array}{cccccc}
            \zetaI & 0 & -1 & 0 \\
            0 & 0 & 1 & \zetaI e^{-\zetaI} \\
            \kr & -1 & 0 & 0 \\
            0 & 1 & 1 & -\kr e^{-\zetaI} \\
        \end{array} \!\!\) \!\!\( \!\!  \begin{array}{c}
                                \MMD{Q}{\TTM}{\I}{\Omega}{\E} \\
                                \MMD{P}{\TTM}{\II}{\Omega}{\L} \\
                                \MMD{Q}{\TTM}{\II}{\Omega}{\L} \\
                                \MMD{P}{\TTM}{\III}{\Omega}{\E} \\
                            \end{array} \!\!\) \= 0$.
\end{tabular}
\end{eqnarray*}
and the global dispersion relations ${D}^{s}_{\E\L\E,\kx,\ky}( \Omega,\xi)=0$ where
\begin{eqnarray}
    \nonumber {D}^{\TTE}_{\E\L\E,\kx,\ky}( \Omega,\xi) &=& \zetaI \(\zetaI + 2 \) e^{-\zetaI}  \\
    \nonumber && \\
    \label{DTM_ELE} {D}^{\TTM}_{\E\L\E,\kx,\ky}( \Omega,\xi) &=& \zetaI \(\zetaI + 2\kr \) e^{-\zetaI} .
\end{eqnarray}
Since $\zetaI>0$, these relations have no real solutions for positive definitive permittivity.
As before, non-trivial TM solutions can arise for media where $\kr(\Omega)$ is not positive definite.

\bibliographystyle{unsrt}
\bibliography{QFIBD}

\end{document}